\documentclass[journal=jacsat,manuscript=article]{achemso}

\SectionNumbersOn
\usepackage[version=3]{mhchem} 
\usepackage{siunitx}      
\usepackage{booktabs}
\usepackage{caption}
\usepackage{subcaption}
\usepackage{graphicx}
\usepackage{tabularx} 
\usepackage[colorlinks=true, linkcolor=blue, citecolor=blue, urlcolor=blue]{hyperref}
\usepackage{mhchem}
\usepackage{siunitx}
\usepackage{makecell} 
\usepackage{adjustbox}
\extrafloats{1000}

\author{Mehedi Hasan}
\affiliation[]{Department of Industrial and Production Engineering, Bangladesh University of Engineering and Technology, Dhaka--1000, Bangladesh}
\email{mehedihasanbuet29@gmail.com}

\author{Khayrul Islam}
\affiliation[]{ Lawrence Livermore National Laboratory, Livermore, California, United States}

\author{Michael T. Kio}
\affiliation[]{Department of Chemical and Biomolecular Engineering, University of Maryland, United States}

\author{AKM Masud}
\affiliation[]{Department of Industrial and Production Engineering, Bangladesh University of Engineering and Technology, Dhaka--1000, Bangladesh}

\title[An \textsf{achemso} demo]
  {Curvature-Dependent Polarity of Interfacial Energy Flow in Functionalized CNT Polymer Nanocomposites: A Reactive Molecular Dynamics Perspective}

\abbreviations{CNT, PDA, PVA}

\begin{document}

\begin{abstract}
    Carbon nanotube (CNT)–polymer composites are widely engineered using surface coatings and chemical treatments to improve interfacial bonding and load transfer. It has been suggested in the nanocomposite literature that nanotube curvature, in conjunction with surface functionalization such as polydopamine (PDA) coating, could serve as an additional control knob for tuning interfacial bonding and energy dissipation in polymer–CNT systems. While experimental and simulation studies have demonstrated the benefits of PDA functionalization, the fundamental mechanism by which nanotube curvature modulates interfacial energy flow and mechanical polarity remains unresolved. This gap is sharpened by a persistent paradox: identical PDA
functionalization strengthens some CNT–polymer systems while weakening others,
a curvature–dependent inconsistency that has remained unexplained.
 Here, we employ reactive molecular dynamics (ReaxFF) simulations to resolve how curvature and PDA functionalization jointly govern interfacial energy evolution in CNT–polyvinyl alcohol (PVA) nanocomposites. Our investigation reveals that curvature and PDA functionalization jointly produce opposite regimes of interfacial energy flow: high-curvature CNTs generate dissipative, frictional interphases, whereas low-curvature CNTs confine energy in rigid, cohesive shells. This polarity inversion originates from a curvature-induced transition in PDA adsorption geometry that transforms the interphase from an energy-releasing to an energy-storing configuration. These results establish curvature as a fundamental design parameter for engineering polymer–nanotube interfaces, offering a predictive route to tune interfacial energy flow, mechanical resilience, and transport properties beyond the limits of conventional chemical functionalization.
\end{abstract}

\section{Introduction}

Polymer--CNT nanocomposites have emerged as lightweight, high performance materials for structural, electronic, and energy applications, combining the exceptional stiffness and conductivity of CNTs with the processability and toughness of polymer matrices~\cite{imtiaz2018review, miyagawa2005mechanical, kausar2021holistic, hasan2024tailoring}. Their potential spans flexible electrodes, sensors, aerospace components, and impact-resistant coatings~\cite{zheng2018high}. In all these systems, the polymer--CNT interface critically governs stress transfer, charge transport, and energy dissipation across multiple length scales. This nanometer-thick interphase often determines whether a composite toughens or fails catastrophically, yet its molecular behavior remains challenging to predict or engineer~\cite{zare2021effects, mohd2021fabrication}.

PVA--CNT nanocomposites have received significant attention due to their complementary attributes: CNTs offer ultrahigh aspect ratios~\cite{falara2022recent}, intrinsic stiffness~\cite{sandhu2007career}, exceptional electrical conductivity~\cite{behabtu2013strong}, and load-bearing capability~\cite{zhou2020role}, while PVA provides excellent processability, flexibility, and hydrogen-bonding interactions~\cite{lekshmi2020recent}. The synergy between the one-dimensional carbon nanostructure and the hydroxyl-rich polymer network underpins their relevance for high-strength fibers, flexible electronics, barrier films, and structural coatings~\cite{yang2015fabrication, yim2025carbon, li2022cnt, liu2023modified}. However, achieving this synergy depends sensitively on nanoscale interfacial interactions.

To enhance CNT--polymer compatibility, numerous surface functionalization strategies have been explored. Oxidative acid treatments introduce carboxyl and hydroxyl groups that increase polarity but may damage the CNT framework~\cite{yang2011functionalization, siwal2020carbon}. Covalent grafting and silanization improve bonding yet risk altering the intrinsic $\pi$-conjugation~\cite{sabet2024innovative, suslova2019effect}. More recently, PDA---a bio-inspired, conformal, and non-destructive coating---has gained attention for its ability to improve adhesion, dispersion, and multifunctionality across polymer matrices~\cite{zeng2025artificial, sharma2025enhancing, wu2025two, jiang2025encapsulating, frenzel2025combined}. PDA wraps CNT surfaces through $\pi$--$\pi$ and hydrogen-bond interactions~\cite{cai2018polydopamine, demirci2021improved}, and first-principles studies show that dopamine adsorbs more strongly on curved CNTs (0.69--0.75~eV) than on flat graphene, indicating curvature-dependent modulation of molecular binding~\cite{kim2020adsorption}. Despite these developments, prior work has predominantly emphasized changes in macroscopic properties---modulus, strength, conductivity---without resolving how surface curvature and molecular geometry influence the interphase formed by such coatings.

A persistent inconsistency in CNT–polymer literature is that identical surface 
functionalization can strengthen one composite while weakening another, an effect 
often observed but seldom mechanistically explained. Prior studies attribute 
improved adhesion to PDA or oxidative functionalities, yet they do not resolve why 
the same chemistry produces opposite mechanical outcomes for different CNT 
chiralities. Motivated by this gap, we hypothesize that nanotube curvature governs 
how functional groups organize and couple with the polymer network, creating 
distinct interfacial architectures on high- versus low-curvature CNTs 
(curvature–functionalization interaction). We further propose that curvature 
controls the direction of interfacial energy exchange, such that high-curvature 
tubes favor dissipative, frictional interphases whereas low-curvature tubes promote 
cohesive, energy-trapping shells (interphase energy-flow polarity). Together, these 
hypotheses frame the central puzzle of this study: why does PDA functionalization 
enhance mobility, porosity, and mechanical resilience in the (10,10) composite but 
yield minimal or adverse effects in the (12,12) counterpart under otherwise 
identical conditions?

To address this curvature-dependent divergence, we employ reactive molecular 
dynamics (ReaxFF) to quantify how curvature and PDA functionalization jointly 
govern interfacial energetics in PVA–CNT nanocomposites. By decomposing bonded, 
nonbonded, and interphase energy contributions during deformation, we determine 
whether curvature induces a reversal in energy-flow polarity—transitioning the 
interphase from dissipative and frictional on high-curvature CNTs to cohesive and 
energy-trapping on low-curvature CNTs. This mechanistic framework establishes 
curvature as a co-equal design parameter with chemistry and provides predictive 
guidelines for tuning toughness, mobility, and dissipation in next-generation 
polymer nanocomposites.

\section{Computational Method}

\subsection{Interatomic Potential}

All molecular dynamics (MD) simulations were performed using the Large‐scale Atomic/Molecular Massively Parallel Simulator (LAMMPS)~\cite{THOMPSON2022108171}. Interatomic interactions were described by the reactive force field (ReaxFF)~\cite{van2001reaxff,chenoweth2008reaxff}, which accounts for both bond formation and dissociation through a continuously updated bond‐order formalism. This potential has been extensively validated for polymer–carbon and organic–inorganic hybrid systems owing to its ability to capture bond rearrangement, charge transfer, and nonbonded interactions within a single framework~\cite{van2001reaxff,sharma2012effect}. 

The total potential energy of the system was expressed as the sum of bonded and nonbonded contributions,
\begin{equation}
E_\mathrm{system}=E_\mathrm{bond}+E_\mathrm{over}+E_\mathrm{under}+E_\mathrm{val}+E_\mathrm{pen}+E_\mathrm{tors}+E_\mathrm{conj}+E_\mathrm{vdW}+E_\mathrm{coulomb},
\end{equation}
where the individual terms correspond to bond stretching, coordination penalties, valence and torsional strains, conjugation, van der Waals, and Coulombic interactions. This formulation enables a seamless transition between covalent and noncovalent regimes, which is essential for describing dynamic polymer–nanotube interfaces.

\subsection{Molecular Model Construction}

The simulation cells consisted of polyvinyl alcohol (PVA) chains interacting with single‐walled carbon nanotubes (SWCNTs) of varying chirality. A 10-mer PVA chain was adopted as the representative polymer segment, consistent with prior studies showing that chains of 5–20 repeat units capture the local adhesion and stress‐transfer behavior governing interfacial mechanics~\cite{rissanou2018atomistic,hasan2024tailoring}. Ten such chains were randomly packed in a periodic tetragonal box (53.4 × 53.4 × 54 Å³) to yield an equilibrium density of 1.26 g cm$^{-3}$~\cite{hashim2014determination}. 

Armchair SWCNTs with chiralities (10,10) and (12,12), corresponding to diameters of 13.56 Å and 16.27 Å respectively, were constructed with a length of 54 Å to match the polymer domain. These chiralities were selected to probe curvature effects on interfacial chemistry and load transfer~\cite{mori2005chirality,natsuki2004effects}. 

To assess the role of surface modification, two functionalization schemes were modeled: nitric‐acid oxidation and polydopamine (PDA) coating. For the HNO$_3$ treatment, intact acid molecules were physisorbed onto the CNT surface to mimic early‐stage oxidation~\cite{ketolainen2018electronic}. For PDA, dopamine monomers were uniformly distributed along the CNT sidewall, reproducing the initial stage of in situ polymerization~\cite{wang2019preparation}. Both treatments preserved the $\pi$-conjugated carbon network while introducing oxygenated and amine groups capable of hydrogen bonding with the PVA matrix. The resulting six models—PVA–CNT, PVA–CNT–HNO$_3$, and PVA–CNT–PDA for each chirality—served as the basis for all simulations.

\subsection{Simulation Protocol}

Each composite was subjected to an identical preparation sequence to ensure structural equilibration and comparability. Geometry optimization was first performed via the conjugate‐gradient algorithm (energy tolerance = $10^{-32}$ kcal mol$^{-1}$, force tolerance = $10^{-6}$ kcal mol$^{-1}$ Å$^{-1}$, 5000 steps). The system was then equilibrated through successive ensembles: NVE (1 ps, heating to 300 K), NPT (5 ps at 300 K, 1 atm), and NVT (5 ps at 300 K) using the Nosé–Hoover thermostat~\cite{plimpton1995fast}. A time step of 0.1 fs was employed to accurately integrate reactive events.

Uniaxial tension was applied along the CNT axis (z-direction) under the NVT ensemble at 300 K, with a constant engineering strain rate of $10^{11}$ s$^{-1}$, consistent with previous high‐strain-rate MD studies~\cite{dupont2012strain}. Although this strain rate ($10^{11}\,\mathrm{s^{-1}}$) is higher than experimental values, it is a standard choice in molecular dynamics simulations to capture relevant deformation behavior within nanosecond timescales. Previous MD studies have shown that lowering the strain rate to the $10^{7}$–$10^{11}\,\mathrm{s^{-1}}$ range primarily influences the post-yield regime without altering the qualitative trends in stiffness and toughness\cite{youssef2024unraveling}.
The resulting stress–strain curves were used to determine Young’s modulus, tensile strength, and toughness.

\subsection{Interphase Identification and RDF-Based Thickness Determination}

The polymer--CNT interphase was defined following established approaches in 
polymer nanocomposite theory, wherein the interphase corresponds to the region of 
modified polymer density and molecular organization surrounding the nanotube 
surface\cite{huang2022interphase}. In molecular simulations, this 
region is commonly identified from deviations in the radial distribution function 
(RDF), which captures how polymer segment density varies with distance from the 
CNT wall. Accordingly, the interphase thickness in all PVA--CNT configurations 
was determined using the center-of-mass RDF between PVA backbone atoms and CNT 
surface carbons.

The RDF was computed as
\begin{equation}
g(r) = \frac{\rho(r)}{\rho_{\mathrm{bulk}}},
\end{equation}
where $\rho(r)$ is the local number density of polymer atoms at a radial 
distance $r$ from the CNT surface and $\rho_{\mathrm{bulk}}$ is the far-field 
polymer density. The boundary of the interphase was defined as the first radial 
position where $g(r)$ decays to unity. This criterion corresponds to the distance beyond the primary 
coordination shell where the polymer density recovers its unperturbed bulk 
behavior, consistent with standard interphase definitions adopted in prior MD 
studies\cite{kumar2023characterizing, su2020mechanical}.

Across all systems, the RDF exhibited a pronounced first maximum associated with 
polymer adsorption at the CNT surface, followed by a monotonic decay. The 
interphase thickness was extracted as the distance from the nanotube surface to 
the position where $g(r) \approx 1.0$, typically spanning $4.5$--$6.0$~\AA\ 
depending on CNT curvature and functionalization. This RDF-based shell was used 
consistently to assign atoms---and their energetic contributions---to either 
the interphase or the bulk polymer domain.

To quantify curvature- and functionalization-dependent interfacial energetics, 
bonded, van der Waals, and electrostatic energy components were computed 
separately for atoms residing within the RDF-defined interphase. This procedure 
enables direct comparison of energy storage, dissipation, and transfer 
mechanisms across CNT chiralities while maintaining a consistent, geometry-
independent definition of the interphase.

\begin{figure*}[htbp]
    \centering
    \includegraphics[width=0.95\textwidth]{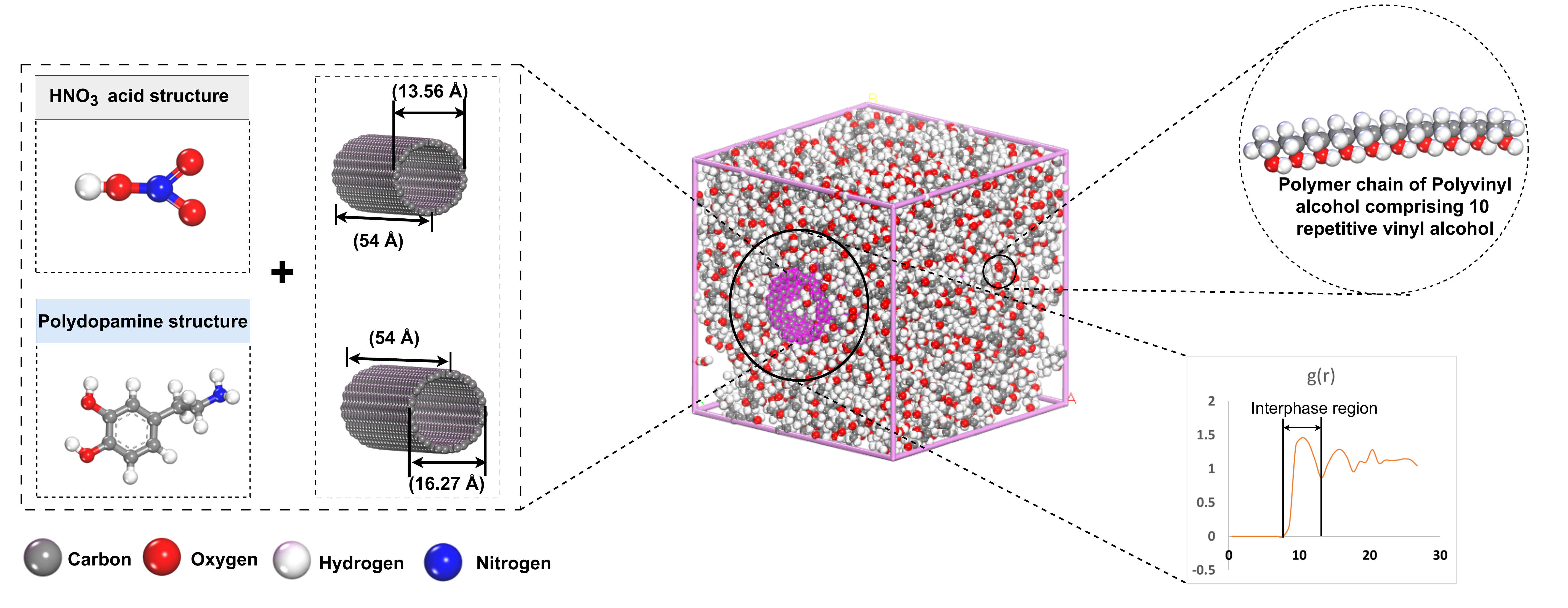}

    \caption{
        \textbf{Schematic illustration of CNT–PVA nanocomposite assembly.}
        Functionalized CNTs (HNO$_3$ and PDA) are embedded within a PVA network,
        showing polymer chain architecture and the interphase region identified
        through radial distribution function (RDF) analysis.
    }
    \label{fig:model_simu}
\end{figure*}

\subsection{Validation of simulation model}

To validate the application of ReaxFF in our research, a benchmark simulation was performed on an armchair (10,10) carbon nanotube (CNT) with a length of 54~\AA{}. The ReaxFF reactive force field was employed to accurately describe carbon–carbon interactions, capturing both bond formation and bond breaking processes through its bond-order formalism. The system underwent energy minimization followed by uniaxial deformation to generate a stress–strain response, from which the Young’s modulus was determined by linear regression of the initial elastic regime. This procedure ensures that the extracted mechanical parameters reflect intrinsic CNT stiffness rather than numerical artifacts from boundary or loading conditions.

The simulated Young’s modulus of the (10,10) CNT (\(E = 631.49~\mathrm{GPa}\)) obtained using the ReaxFF potential demonstrates excellent agreement with previously reported theoretical and experimental values (Table~\ref{tab:youngs_modulus_comparison}). Experimental investigations by Salvetat \textit{et~al.}\cite{salvetat1999elastic} and Yu \textit{et~al.}\cite{yu2000strength} reported moduli ranging from \(800 \pm 410~\mathrm{GPa}\) to \(270{-}950~\mathrm{GPa}\), respectively, while atomistic simulations employing the COMPASS,\cite{sharma2013molecular} REBO,\cite{wenxing2004simulation} and Morse\cite{lu2005effect} potentials yielded values of 519, $935.805 \pm 0.618$, and $840 \pm 20$, respectively. The present ReaxFF result thus lies well within the experimentally observed and computationally predicted window for CNTs, validating the mechanical fidelity of the potential for sp\textsuperscript{2}-bonded carbon frameworks.

The simulated Young’s modulus of the (10,10) CNT (631~GPa) agrees well with reported atomistic results for the same chirality (519~GPa~\cite{sharma2013molecular}) and lies near the average modulus of armchair CNTs (\(\approx 602~\mathrm{GPa}\)~\cite{sharma2013molecular}), confirming the reliability of the ReaxFF description of CNT elasticity. This consistency across independent experimental and computational benchmarks reinforces the applicability of ReaxFF for predicting nanoscale mechanical behavior.

\begin{table}[htbp]
\centering
\caption{Analyzing Young’s Modulus of Carbon Nanotubes in Comparison with Existing Literature Values}
\begin{tabular}{l c}
\hline
\textbf{Investigations} & \textbf{Young’s Modulus (GPa)} \\
\hline
Present work & 631.49 \\
Salvetat et al.\ [experimental]\cite{salvetat1999elastic} & $800 \pm 410$ \\
Yu et al.\ [experimental]\cite{yu2000strength} & 270 -- 950 \\
Sharma et al.\ [COMPASS potential]\cite{sharma2013molecular} & 519 \\
WenXing et al.\ [REBO and LJ potential]\cite{wenxing2004simulation} & $935.805 \pm 0.618$ \\
Qiang et al.\ [Morse potential]\cite{lu2005effect} & $840 \pm 20$ \\
\hline
\end{tabular}
\label{tab:youngs_modulus_comparison}

\end{table}

This comprehensive approach enables the nuanced simulation of PVA–CNT nanocomposites, capturing the essence of their mechanical performance under diverse loading and environmental conditions. The use of the ReaxFF force field, with its detailed accounting of bond, angle, torsion, conjugation, and nonbonded energy contributions, underscores the sophistication and precision of current atomistic methodologies in elucidating the molecular foundations of material properties. The validated ReaxFF framework thus provides a robust foundation for exploring curvature-, chemistry-, and interphase-dependent mechanics in functionalized CNT–polymer nanocomposites.

\section{Results and Discussion}

\subsection{mechanical, trnasportation, and structal properites analysis}

\subsubsection{Structural Integrity Analysis}

Figure~\ref{fig:mechanical_data} (a, b) presents the stress--strain responses of PVA--CNT nanocomposites subjected to uniaxial tension at 300~K. In all cases, the initial regime (up to $\approx$~0.18 strain) exhibits a linear elastic response, characteristic of efficient load transfer across the polymer--nanotube interface. For the PVA--CNT(10,10) systems functionalized with \ce{HNO3} and polydopamine (PD), the stress subsequently rises nonlinearly, reaching markedly higher ultimate strengths before abrupt failure---reflecting enhanced interfacial adhesion and improved stress transfer to the CNT backbone. In contrast, the pristine PVA--CNT(10,10) composite and all (12,12) systems, regardless of surface treatment, display a more gradual increase in stress after the elastic region, culminating at lower ultimate stresses followed by a sharp drop to near-zero stress, indicative of CNT fracture and interfacial debonding. These trends highlight the distinct reinforcement mechanisms governed by nanotube chirality and surface functionalization.

\medskip

The mechanical properties, derived from these stress--strain profiles, are summarized in Table~\ref{tab:actual_mechanical_properties}. The Young’s modulus for each composite was obtained  through nonlinear regression
analysis\cite{brown2001step}. For the PVA--CNT(10,10) systems, the pristine composite exhibits a Young’s modulus of 53.36~GPa and tensile strength of 9.91~GPa, reflecting a relatively rigid yet brittle behavior with limited toughness 1.25~GJ\,m$^{-3}$. After \ce{HNO3} treatment, oxygenated surface groups slightly reduce the stiffness 49.97~GPa but significantly improve load transfer and energy absorption, doubling the toughness to 2.20~GJ\,m$^{-3}$. For the PD-functionalized composite that results in simultaneous increases in modulus 58.79~GPa, strength 12.22~GPa, and toughness 2.61~GJ\,m$^{-3}$. These results indicate that PD functionalization effectively overcomes the classical stiffness--toughness trade-off by forming a cohesive and energy-dissipative interphase that supports extensive strain delocalization, consistent with the stress--strain behavior shown in Figure~\ref{fig:mechanical_data} (a, b).

\medskip

For the PVA--CNT(12,12) systems, the effect of chirality manifests as higher stiffness but reduced deformability. The pristine composite attains the highest modulus 64.79~GPa and 11.41~GPa, accompanied by limited toughness {1.28~GJ\,m$^{-3}$. After \ce{HNO3} treatment, mild oxidation lowers the modulus slightly to 60.01~GPa while enhancing both strength 11.70~GPa and toughness 1.43~GJ\,m$^{-3}$, reflecting improved interfacial compatibility and partial defect-induced bonding. PD functionalization maintains a comparable modulus 61.77~GPa and strength 11.80~GPa, while preserving moderate toughness 1.32~GJ\,m$^{-3}$. These results suggest that for lower-curvature CNTs, PD provides stable interfacial load pathways without compromising stiffness, achieving a balanced mechanical performance. Consequently, chirality plays a key role in defining the intrinsic stiffness baseline, while surface functionalization modulates the interfacial load-transfer efficiency and overall mechanical response of the composite, as illustrated in Figure~\ref{fig:mechanical_data} (a, b).

\medskip

The simulated strengthening trends are in excellent agreement with our earlier experimental observations reported by Moon \textit{et~al.}\cite{moon2025synthesis}, where A.~K.~M.~Masud and co-workers examined PVA films reinforced with pristine, acid-treated, and polydopamine-modified multiwalled CNTs. Although explicit numerical values for Young’s modulus and tensile strength were not provided, the reported stress--strain curves exhibited the same progressive enhancement sequence identified in our simulations: PVA/CNT~$<$~PVA/CNT--HNO$_3$~$<$~PVA/CNT--PDA. In that study, the PDA-coated composites displayed the steepest initial elastic slope and the highest ultimate stress, signifying markedly improved interfacial adhesion and load-transfer efficiency. This strong correspondence between our previous experimental findings and the present atomistic predictions validates the ReaxFF framework and highlights the universal role of curvature- and chemistry-driven interphase strengthening in CNT--polymer nanocomposites.

\subsubsection{Energy Dissipation and Mechanical Integrity Under Tensile Strain}

Figure~\ref{fig:mechanical_data}c--d depict the potential energy evolution of PVA--CNT nanocomposites under uniaxial tension. The total energy decreases monotonically during equilibration and subsequently rises with strain, reflecting the progressive stretching of polymer chains and nanotube frameworks prior to failure. The magnitude and shape of these energy--strain curves are strongly dependent on both CNT chirality and surface functionalization.

\medskip

For the PVA--CNT(10,10) systems (Figure~\ref{fig:mechanical_data}c), the pristine composite exhibits a steep increase in potential energy up to approximately 0.3 strain, consistent with elastic bond stretching followed by interfacial slippage. \ce{HNO3}-functionalized composites display a similar trend but reach a higher peak energy ($-1.94\times10^{6}$~kcal\,mol$^{-1}$), suggesting enhanced energy absorption through localized bond rearrangement and moderate interfacial degradation. In contrast, the PD-functionalized composite maintains a consistently less negative energy baseline ($\sim-1.53\times10^{6}$~kcal\,mol$^{-1}$) and a smoother strain-dependent rise, indicating more efficient load redistribution and delayed onset of structural instability. This stabilized energetic response corroborates the superior toughness observed in Table~\ref{tab:actual_mechanical_properties}.

\medskip

A similar yet chirality-dependent behavior is observed for the PVA--CNT(12,12) systems (Figure~\ref{fig:mechanical_data}d). The pristine composite begins near $-1.99\times10^{6}$~kcal\,mol$^{-1}$ and exhibits a steep energy rise until $\sim$0.20 strain, followed by a sudden drop corresponding to the onset of fracture and interfacial debonding. The \ce{HNO3}-treated sample shows a comparable but slightly earlier energy release, consistent with the presence of acid-induced defect sites that facilitate premature local failure. By contrast, the PD-functionalized (12,12) composite displays a more moderate energy increase and a delayed energy release event, implying enhanced adhesion and controlled interfacial energy dissipation. 

\medskip

These observations collectively demonstrate that both chirality and surface functionalization critically determine how mechanical work is stored and released during deformation. The energetic response thus provides a direct atomistic signature of interfacial integrity, complementing the macroscopic mechanical trends discussed above.

\subsubsection{Diffusion behavior and mean-squared displacement analysis}

The diffusivity of all PVA--CNT nanocomposites is shown in Figure~\ref{fig:mechanical_data}g, providing molecular-level insight into thermally activated segmental dynamics. The ensemble-averaged MSD at time $t$ is defined as~\cite{wunderle2009molecular}:

\[
\text{MSD}(t - t_0) = \frac{1}{6N} \sum_{i=1}^{N} \left[ \mathbf{r}_i(t) - \mathbf{r}_i(t_0) \right]^2 \tag{1}
\]

where $\mathbf{r}_i(t)$ denotes the position of the $i^{\text{th}}$ atom at time $t$, and $N$ is the total number of atoms. The diffusion coefficient $D$ was evaluated from Einstein’s relation~\cite{boyd1991molecular}:

\[
D = \frac{1}{6N} \lim_{t \to \infty} \frac{d}{dt} \sum_{i=1}^{N} \left[ \mathbf{r}_i(t) - \mathbf{r}_i(0) \right]^2 \tag{2}
\]

The linear regime of each MSD curve was fitted by least-squares regression~\cite{chipman1979efficiency} using
\[
\text{MSD}(t) = a\,t + b,
\]
from which $D$ was obtained as~
\[
D = \frac{a}{2d},
\]
where $d = 3$ denotes the spatial dimensionality of the system. All MSD values were computed using the \texttt{compute msd\_all all msd} command in LAMMPS under the NVT ensemble, capturing collective segmental displacements within the polymer matrix. No solvent molecules were included; therefore, the derived diffusion coefficients represent effective nanoscale mobilities rather than classical self-diffusion constants. Complete simulation parameters are provided in the Supplementary Information.

\medskip

The diffusion coefficients exhibit a clear dependence on both chirality and surface chemistry. For pristine composites, PVA--CNT(10,10) shows higher mobility than PVA--CNT(12,12) \big($3.31\times10^{-6}$ vs.\ $1.89\times10^{-6}$~\(\mathrm{cm^2\,s^{-1}}\)\big), indicating that the higher curvature of (10,10) introduces additional free volume and segmental freedom. \ce{HNO3} functionalization amplifies this chirality contrast, yielding $4.31\times10^{-6}$ for (10,10) but only $8.86\times10^{-7}$ for (12,12), consistent with acid-induced interfacial disorder that loosens local packing for the high-curvature tubes while densifying the lower-curvature interface. Polydopamine (PD) produces the highest mobility for (10,10) \big($4.65\times10^{-6}$\big) yet a moderate value for (12,12) \big($1.67\times10^{-6}$\big), placing PD-treated (12,12) below its pristine counterpart and well above the acid-treated system.

\subsubsection{Porosity evolution and interphase stability}

To further elucidate the structural integrity of the PVA--CNT nanocomposites under tensile loading, the evolution of voids  ware quantified through porosity analysis (Figure~\ref{fig:mechanical_data}f--g). The porosity (\%) at each strain state was calculated as:
\[
\text{Porosity} = \frac{V_\text{total} - V_\text{solid}}{V_\text{total}} \times 100 \tag{3}
\]
where $V_\text{total}$ is the instantaneous simulation box volume and $V_\text{solid}$ represents the occupied atomic volume determined using a probe radius of 4~\AA. This metric captures the volumetric void fraction generated during deformation and serves as a mesoscale descriptor of interfacial stability.

\medskip

For the PVA--CNT(10,10) systems (Figure~\ref{fig:stat_visu}f), the porosity remains negligible below $\varepsilon \approx 0.05$, confirming that deformation is primarily elastic with no microvoid formation. Beyond this strain, porosity increases gradually until $\varepsilon \approx 0.15$, after which a pronounced rise occurs. The pristine composite reaches a porosity of $\sim$19\% at $\varepsilon=0.30$, while the \ce{HNO3}-treated system attains a slightly lower value ($\sim$18.15\%). In contrast, the PD-functionalized composite exhibits a steady and significantly higher porosity, reaching $\sim$27\% at $\varepsilon=0.30$. The higher porosity observed in the PD-functionalized (10,10) system is accompanied by a smooth potential-energy evolution and the highest toughness, indicating progressive rather than abrupt structural deformation. This behavior suggests that void formation occurs through distributed interfacial yielding and chain rearrangement—consistent with a ductile, energy-dissipative interphase rather than premature damage.

\begin{table}[htbp]
\centering
\caption{Mechanical and physical properties of the developed nanocomposites.}
\label{tab:actual_mechanical_properties}

\small   
\setlength{\tabcolsep}{4pt}  

\begin{tabular}{l l c c c c c c}
\toprule
\textbf{Chirality} & \textbf{Treatment} 
& \makecell{\textbf{Modulus}\\(GPa)}
& \makecell{\textbf{Strength}\\(GPa)}
& \makecell{\textbf{Toughness}\\(GJ/m$^{3}$)}
& \makecell{\textbf{Energy}\\(kcal/mol)}
& \makecell{\textbf{Diffusivity}\\(cm$^{2}$/s)}
& \makecell{\textbf{Porosity}\\(\%)} \\
\midrule

\makecell[l]{(10,10)\\13.56 Å} & Non-treated 
& 53.36 & 9.91 & 1.25 
& -1.96627$\times 10^{6}$ & 3.31$\times 10^{-6}$ & 19.03 \\

\makecell[l]{(12,12)\\16.27 Å} & Non-treated 
& 64.79 & 11.41 & 1.28 
& -1.96363$\times 10^{6}$ & 1.89$\times 10^{-6}$ & 17.77 \\

\makecell[l]{(10,10)\\13.56 Å} & HNO\textsubscript{3}-treated 
& 49.97 & 10.08 & 2.20 
& -1.94068$\times 10^{6}$ & 4.31$\times 10^{-6}$ & 18.15 \\

\makecell[l]{(12,12)\\16.27 Å} & HNO\textsubscript{3}-treated 
& 60.01 & 11.70 & 1.43 
& -1.96135$\times 10^{6}$ & 8.86$\times 10^{-7}$ & 18.58 \\

\makecell[l]{(10,10)\\13.56 Å} & PD-treated 
& 58.79 & 12.22 & 2.61 
& -1.53453$\times 10^{6}$ & 4.65$\times 10^{-6}$ & 26.67 \\

\makecell[l]{(12,12)\\16.27 Å} & PD-treated 
& 61.77 & 11.80 & 1.32 
& -1.96474$\times 10^{6}$ & 1.67$\times 10^{-6}$ & 15.09 \\
\bottomrule
\end{tabular}
\end{table}


\begin{figure}[htbp]
  \centering
  \includegraphics[width=\textwidth]{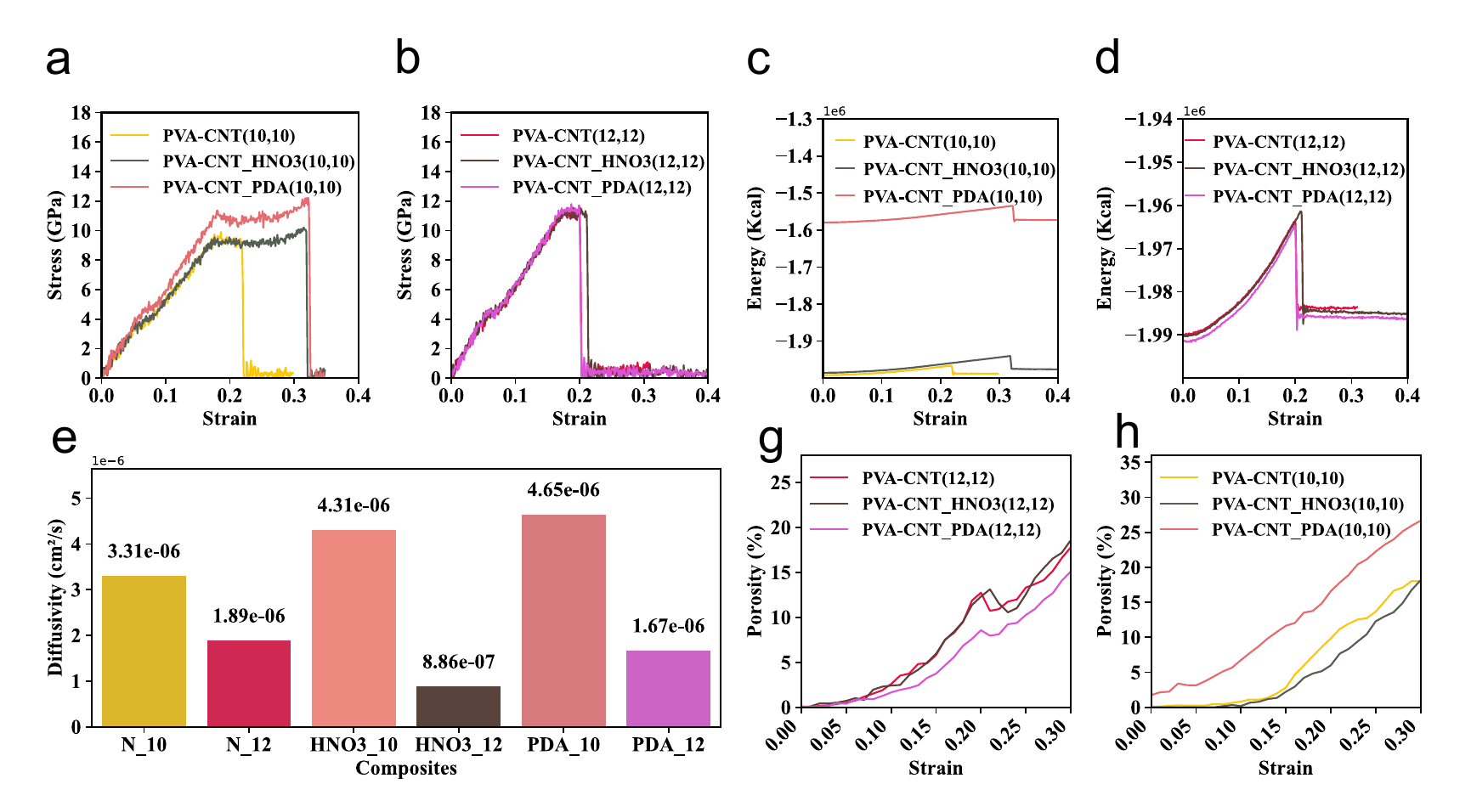}
  \caption{Mechanical and transport properties of PVA–CNT nanocomposites. 
(a, b) Stress–strain curves for (10,10) and (12,12) CNT-based composites with pristine, HNO$_3$-treated, and polydopamine (PD)-functionalized CNTs. 
(c, d) Energy–strain responses for (10,10) and (12,12) systems. 
(e) Diffusivity values of different composites. 
(f, g) Porosity evolution with strain for (10,10) and (12,12) composites, respectively.}

  \label{fig:mechanical_data}
\end{figure}

\newpage

\subsection{Structure–Property Trends and Curvature-Dependent Deviations}

\subsubsection{Canonical Scaling Laws Linking Porosity, Mechanics, and Transport}

Porosity strongly influences the mechanical and transport behavior of polymers and porous solids. Increasing porosity ($\phi$) generally reduces stiffness, strength, and toughness due to stress concentration and reduced load-bearing area~\cite{Gibson1997,Rice1968,Huang2020}. These trends are classically captured by the Gibson–Ashby relation:
\[
    \frac{E}{E_s} \approx (1-\phi)^n,\qquad 
    \frac{\sigma}{\sigma_s} \approx (1-\phi)^m,
\]
with $n,m\approx2$ for open-cell materials~\cite{Gibson1997}. Toughness similarly declines as pores facilitate early crack initiation~\cite{Anderson2017,Budiansky1978}.

Transport properties typically show the opposite trend. Increased porosity reduces tortuosity and enhances effective diffusivity~\cite{Millington1961,Shen2018}. In polymer–CNT systems, the mean squared displacement (MSD) further reflects local chain mobility~\cite{skountzos2018molecular}, which may either support energy dissipation in well-bonded regions~\cite{Bucknall2013,Arruda1993} or indicate structural weakness in poorly adhered interfaces~\cite{Qi2009}. These canonical behaviors provide a reference for interpreting the simulated CNT–polymer systems.

\subsubsection{Mechanistic Analysis of Structure–Property Relationships}

\paragraph{Qualitative Trends Relative to Canonical Scaling.}
Across the six CNT–PVA systems, the scatter plots in Fig.~\ref{fig:scatter_matrix} largely reflect expected canonical trends: lower porosity is associated with higher modulus and strength, and moderate mobility accompanies moderate toughness. One configuration, however, departs from this pattern. The PDA-functionalized (10,10) system occupies a distinct region of the property space, combining elevated porosity and mobility with enhanced mechanical performance. In contrast, all (12,12) systems cluster along the conventional scaling behavior. These qualitative deviations suggest that an additional geometric factor influences the observed responses.

\paragraph{Curvature-Dependent Grouping.}
Further examination of stress–strain curves, porosity evolution, and MSD profiles (Fig.~\ref{fig:mechanical_data}) reveals a clear curvature-based grouping. The three (12,12) systems exhibit cohesive mechanical behavior with limited porosity growth and restricted chain mobility. In contrast, the (10,10) systems—particularly the PDA- and HNO$_3$-modified variants—show pronounced mobility and porosity changes, indicating a different mode of interphase evolution. Because chemical treatments are identical across chiralities, this consistent separation highlights curvature as a key organizing parameter governing the structural response.

\paragraph{Percent-Change Contrasts and Chirality–Functionalization Interaction.}
Percent-change contrasts (Table~\ref{tab:contrasts_all_features}) further emphasize these differences. At (10,10), PDA functionalization increases toughness, diffusivity, and porosity while also enhancing strength. HNO$_3$ treatment produces similar trends. At (12,12), however, the same treatments yield modest improvements or even reductions in mechanical performance. These opposite directional responses indicate a strong curvature–functionalization interaction rather than a uniform additive effect.

\paragraph{Interaction Polarity.}
The polarity of these responses becomes apparent when the direction of property changes is compared across chiralities. Features that increase at (10,10) frequently decrease at (12,12) under identical chemical modification and loading conditions. Such mirrored responses are not accounted for by classical porosity–mechanics or mobility–toughness relations, suggesting that curvature alters the fundamental mode by which the interphase accommodates functionalization.

\begin{figure*}[htbp] \centering \includegraphics[width=\linewidth]{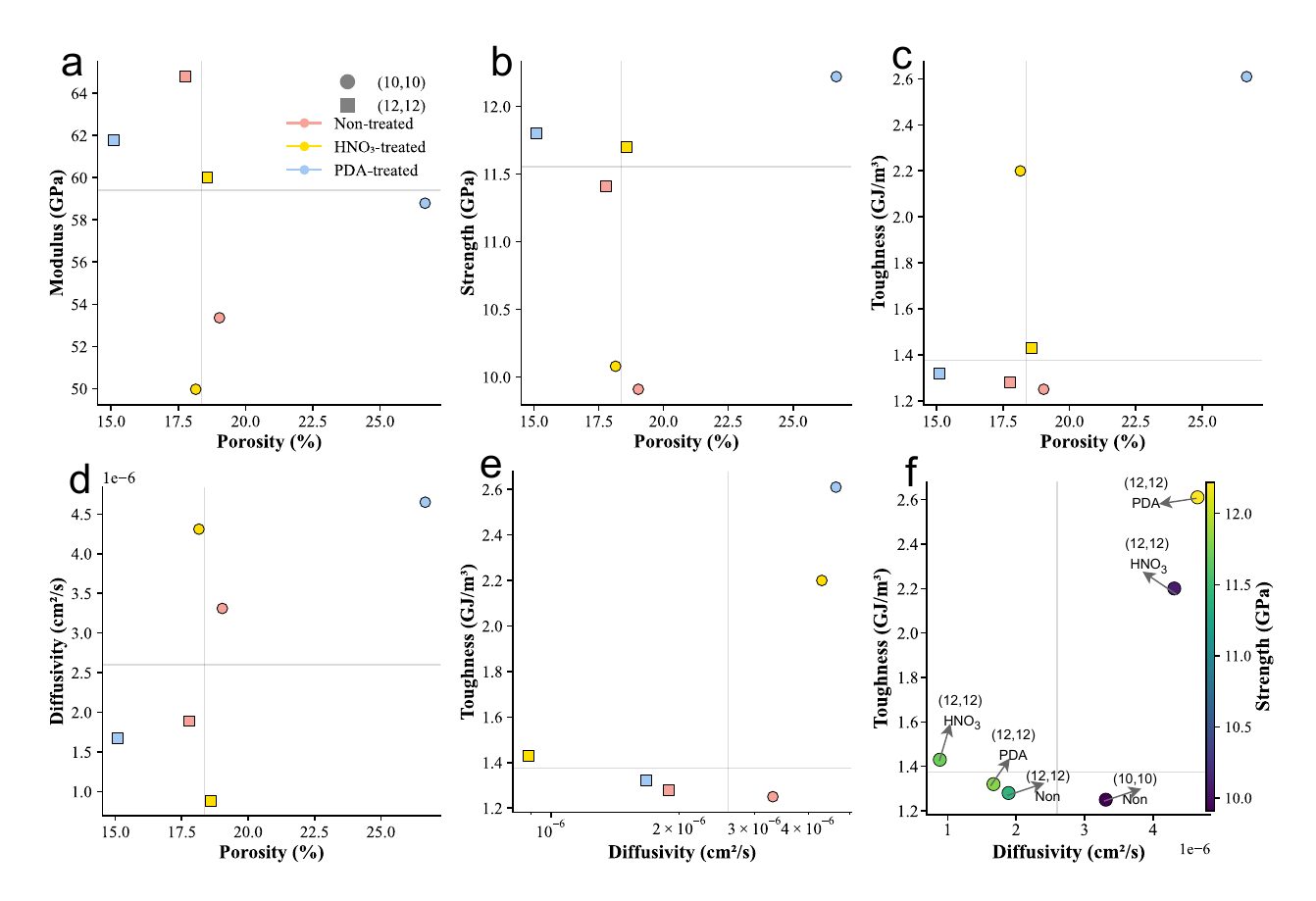} \caption{ Correlation plots among key structural, transport, and mechanical descriptors for all simulated CNT–polymer composites. Panels (a–c) show modulus, strength, and toughness as functions of porosity, respectively; panel (d) presents diffusivity versus porosity; panel (e) correlates toughness with diffusivity; and panel (f) combines diffusivity, toughness, and strength into a unified mechano–transport map. Each data point corresponds to a distinct CNT chirality \((10,10)\) or \((12,12)\) under different surface conditions (non-treated, HNO$_3$-treated, and PDA-treated).} \label{fig:scatter_matrix} \end{figure*} \begin{table}[htbp] \centering \caption{Percent-change contrasts (\%) across all features. \small $\displaystyle \Delta\%=\frac{\text{after}-\text{before}}{\text{before}}\times100\%$; for “X vs Non at Y”: after$=$X\_Y, before$=$Non\_Y; for “curvature (12,12 vs 10,10)$:$ after$=$Feature\_{12,12}, before$=$Feature\_{10,10}. Positive = increase relative to the denominator.} \label{tab:contrasts_all_features} \resizebox{\textwidth}{!}{ \begin{tabular}{lrrrrrr} \toprule Contrast & Modulus (GPa) & Strength (GPa) & Toughness (GJ/m\textsuperscript{3}) & Energy (kcal/mol) & MSD (cm\textsuperscript{2}/s) & Porosity (\%) \\ \midrule PD vs Non at (10,10) & 10.18 & 23.31 & 108.80 & -21.96 & 40.48 & 40.15 \\ PD vs Non at (12,12) & -4.66 & 3.42 & 3.13 & 0.06 & -11.64 & -15.08 \\ HNO\textsubscript{3} vs Non at (10,10) & -6.35 & 1.72 & 76.00 & -1.30 & 30.21 & -4.62 \\ HNO\textsubscript{3} vs Non at (12,12) & -7.38 & 2.54 & 11.72 & -0.12 & -53.12 & 4.56 \\ PD curvature (12,12 vs 10,10) & 5.07 & -3.44 & -49.43 & 28.04 & -64.09 & -43.42 \\ HNO\textsubscript{3} curvature (12,12 vs 10,10) & 20.09 & 16.07 & -35.00 & 1.07 & -79.44 & 2.37 \\ \bottomrule \end{tabular}} \end{table} \newpage

\subsubsection{Curvature-Dependent Functionalization Puzzle}

Together, these observations define a central question.  
PDA functionalization substantially enhances mobility, porosity, and mechanical performance in the high-curvature (10,10) composite, yet produces only modest or adverse effects in the low-curvature (12,12) composite. Thus, identical chemistry generates two contrasting interphase behaviors under otherwise comparable conditions.

\begin{quote}
\textbf{Puzzle:}  
Why does PDA functionalization reinforce the (10,10) CNT–PVA composite—producing simultaneous increases in mobility, porosity, and toughness—while the (12,12) composite remains comparatively constrained and weakly responsive to the same treatment?
\end{quote}

This curvature-dependent divergence motivates the energetic and interphase-level analyses presented in the following sections.

\subsection{Energetic Resolution of the Curvature–Functionalization Puzzle}

\paragraph{Potential-Energy Evolution of the Composite and Its Components.}
The total potential energy profiles (\autoref{fig:energy_evolution}a–d) exhibit a distinct curvature-dependent contrast between the (10,10) and (12,12) PDA–PVA–CNT composites. For the high-curvature (10,10) system, the total potential energy ($E_\mathrm{total}$) increases smoothly from $-1.579\times10^{6}$ to $-1.539\times10^{6}$~kcal~mol$^{-1}$ as strain rises from 0 to 30\%, corresponding to an overall energy gain of $\sim4.0\times10^{4}$~kcal~mol$^{-1}$ (2.5\% of the initial value). Decomposition of this total energy reveals that the CNT potential energy ($E_\mathrm{CNT}$) rises from $-1.77\times10^{5}$ to $-1.37\times10^{5}$~kcal~mol$^{-1}$, the PVA energy ($E_\mathrm{polymer}$) changes marginally from $-1.399\times10^{6}$ to $-1.398\times10^{6}$~kcal~mol$^{-1}$, and the PDA component remains nearly constant at $E_\mathrm{PDA}\approx -2.31\times10^{3}$~kcal~mol$^{-1}$. This steady, monotonic increase implies that mechanical work is uniformly distributed and gradually accommodated throughout the network, with strain energy being dissipated across both the CNT and polymer backbones. In contrast, the low-curvature (12,12) composite follows a non-monotonic trajectory: $E_\mathrm{total}$ decreases from $-1.991\times10^{6}$ to $-1.965\times10^{6}$~kcal~mol$^{-1}$ up to $\varepsilon\approx0.20$, after which a sudden energy jump occurs, marking the onset of interfacial rupture. The CNT contribution ($E_\mathrm{CNT}$) increases from $-2.12\times10^{5}$ to $-1.88\times10^{5}$~kcal~mol$^{-1}$, the polymer energy changes slightly from $-1.775\times10^{6}$ to $-1.773\times10^{6}$~kcal~mol$^{-1}$, and the PDA energy fluctuates between $-2.02\times10^{3}$ and $-1.94\times10^{3}$~kcal~mol$^{-1}$, suggesting unstable interfacial adhesion prior to failure. Together, these trends confirm that the (10,10) system stores and dissipates strain energy in a continuous, distributed fashion through its curved interphase, while the flatter (12,12) geometry localizes energy at the CNT–PDA junction, resulting in a cohesive energy trap that releases abruptly under tensile deformation.

\paragraph{van der Waals Energy Characteristics.}
The van der Waals (vdW) interaction energy (\autoref{fig:energy_evolution}e–h) reveals curvature-dependent but quantitatively comparable behaviors in the (10,10) and (12,12) PDA–PVA–CNT composites. For the (10,10) configuration, the total vdW energy ($E_\mathrm{vdW,\,total}$) decreases slightly from $6.30\times10^{5}$ to $6.27\times10^{5}$~kcal~mol$^{-1}$ over 0–30\% strain, showing mild oscillations of about $\pm(2$–$3)\!\times\!10^{3}$~kcal~mol$^{-1}$. The CNT contribution declines from $5.94\times10^{4}$ to $5.55\times10^{4}$~kcal~mol$^{-1}$, whereas the polymer term remains nearly constant at $\sim5.70\times10^{5}$~kcal~mol$^{-1}$. The PDA vdW energy fluctuates within $960$–$1015$~kcal~mol$^{-1}$ (average $978.5$~kcal~mol$^{-1}$), showing modest, continuous readjustment during deformation. Together, these data indicate that the (10,10) interface undergoes frequent, low-amplitude fluctuations that correspond to small-scale rearrangements of $\pi$–$\pi$ and hydrogen-bond contacts, but without large energy release or loss of adhesion. In contrast, the (12,12) composite maintains a similar overall magnitude of vdW energy—nearly constant between $7.90\times10^{5}$ and $7.96\times10^{5}$~kcal~mol$^{-1}$—through most of the deformation range, followed by a gentle decrease of $\sim2\times10^{3}$~kcal~mol$^{-1}$ near 20\% strain. The CNT and polymer components remain close to $7.0\times10^{4}$ and $7.23\times10^{5}$~kcal~mol$^{-1}$, respectively, while the PDA vdW term varies slightly between $860$ and $920$~kcal~mol$^{-1}$ (average $890.8$~kcal~mol$^{-1}$). Quantitatively, both systems exhibit similar fractional fluctuation amplitudes (about 0.5–0.6\% of their mean energy), but differ in temporal character: the (10,10) profile shows continuous small oscillations, whereas the (12,12) curve remains smooth until a single late relaxation. These observations confirm that curvature does not strongly alter the absolute magnitude of vdW cohesion but changes how that cohesion evolves with strain—frequent incremental adjustments for the more curved (10,10) tube versus a steadier, delayed relaxation for the flatter (12,12) surface.

\paragraph{Electrostatic-Energy Response.}
The electrostatic-energy profiles (\autoref{fig:energy_evolution}i–l) show that both curvature regimes maintain overall stability, with only small but measurable differences in magnitude and temporal variation. In the (10,10) composite, the total electrostatic energy ($E_\mathrm{elec,\,total}$) changes slightly from $-1.71\times10^{5}$ to $-1.70\times10^{5}$~kcal~mol$^{-1}$ as strain increases from 0 to 30\%, producing a shallow rise of roughly $0.6$\%. This trend is accompanied by low-amplitude oscillations of $\sim5\!\times\!10^{2}$–$1\!\times\!10^{3}$~kcal~mol$^{-1}$, mainly originating from the polymer phase ($E_\mathrm{elec,\,polymer}\!\approx\!-1.69\times10^{5}$~kcal~mol$^{-1}$). The PDA electrostatic energy varies slightly between $-305$ and $-270$~kcal~mol$^{-1}$, showing weak fluctuations that parallel the minor vdW serrations, while the CNT contribution remains near zero. These small but correlated variations suggest mild polarization changes within the interphase but no significant charge redistribution or structural instability. In comparison, the (12,12) composite maintains a nearly flat electrostatic profile, with $E_\mathrm{elec,\,total}$ ranging narrowly from $-2.19\times10^{5}$ to $-2.15\times10^{5}$~kcal~mol$^{-1}$ through the full deformation range. The polymer term dominates this contribution ($E_\mathrm{elec,\,polymer}\!\approx\!-2.15\times10^{5}$~kcal~mol$^{-1}$), and the PDA component remains stable between $-200$ and $-180$~kcal~mol$^{-1}$ without systematic variation. The corresponding curves (\autoref{fig:energy_evolution}k,l) confirm smooth, monotonic behavior for both systems, with the (10,10) trace exhibiting minor oscillations and the (12,12) trace remaining almost linear. Quantitatively, the total energy difference between the two systems is dominated by curvature-related contact area rather than dynamic effects. Overall, the electrostatic energies indicate that both interfaces preserve strong and stable dipolar interactions under strain, with the (10,10) system showing only slightly greater sensitivity to deformation while the (12,12) system remains nearly invariant.

\begin{figure*}[htbp]
    \centering
    \includegraphics[width=0.95\textwidth]{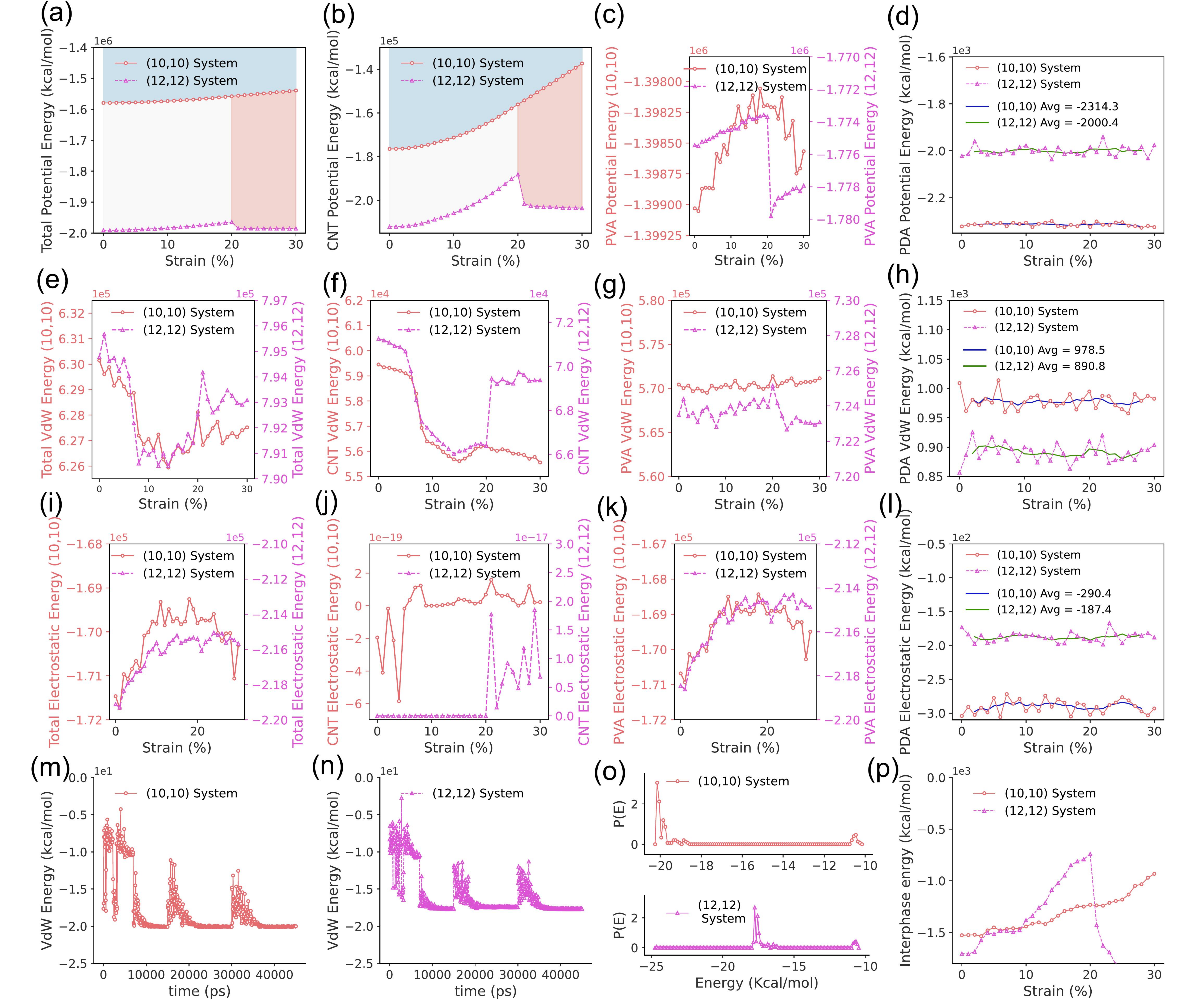}

    \caption{
        \textbf{Energetic resolution of the curvature–functionalization puzzle.}
        (a–d) Total, CNT, PVA, and PDA potential energies versus strain. 
        (e–h) van der Waals (vdW) energy evolution for each subsystem. 
        (i–l) Electrostatic-energy trends showing curvature-dependent polarization. 
        (m–n) Time-resolved vdW-energy fluctuations distinguishing sliding vs rupture. 
        (o) PDA adsorption energy-distribution profile. 
        (p) Interphase-energy evolution showing polarity inversion between (10,10) and (12,12). 
        All energies are expressed in kcal~mol$^{-1}$.
    }
    \label{fig:energy_evolution}
\end{figure*}

\paragraph{Interphase–Energy Evolution and Root Causes of the Puzzle.}
The curvature‐dependent polarity observed during mechanical loading originates fundamentally from the PDA–CNT adsorption stage, where curvature modulates both the depth and heterogeneity of the van der Waals potential wells. The time‐resolved $E_\mathrm{vdW}$ trajectories (\autoref{fig:energy_evolution}m–n) show that PDA adsorption on the high‐curvature (10,10) surface stabilizes faster and into a slightly deeper potential minimum ($E_\mathrm{vdW}\!\approx\!-2.0\times10^{1}$~kcal~mol$^{-1}$) than on the low‐curvature (12,12) surface ($E_\mathrm{vdW}\!\approx\!-1.77\times10^{1}$~kcal~mol$^{-1}$). From (\autoref{fig:energy_evolution}o), the (10,10) system, the narrow, sharply defined energy well and rapid damping of fluctuations indicate localized anchoring at discrete curved sites, where $\pi$–$\pi$ overlap is strong but spatially heterogeneous—allowing partial chain mobility and frictional readjustment. Conversely, in the (12,12) system, the broader, multi‐peak distribution and slower relaxation reflect more uniform but weaker adsorption across a flatter surface, where PDA chains adopt extended $\pi$–$\pi$ registry and reduced configurational freedom. Thus, curvature predefines the energetic landscape: the (10,10) interface forms a mosaic of deep, isolated potential basins permitting reversible motion, whereas the (12,12) interface forms a continuous adhesive plateau that restricts local reconfiguration.

To examine how this adsorption contrast manifests during deformation, we track the evolution of the \emph{intercomponent interaction energy} across the CNT–PDA–PVA triad. In this work, the interaction energy was computed as
\[
E_\mathrm{interphase} = E_\mathrm{total} - (E_\mathrm{CNT} + E_\mathrm{PDA} + E_\mathrm{PVA}),
\]
which isolates the net van der Waals and electrostatic interactions between components.\footnote{A spatially defined interphase  (RDF-based shell) is also analyzed separately in Section~2; here, $E_\mathrm{interphase}$ is used solely to quantify how curvature alters the direction of intercomponent energy exchange.}

 For the (10,10) system, $E_\mathrm{interphase}$ increases steadily from $-1.53\times10^{3}$ to $-9.3\times10^{2}$~kcal~mol$^{-1}$ as strain rises to 30\%, signifying gradual energy release and redistribution into the polymer matrix. The positive $\mathrm{d}E_\mathrm{interphase}/\mathrm{d}\varepsilon$ reflects a continuously relaxing, dissipative interface where nanoscale slip and reversible $\pi$–$\pi$ torsion accommodate deformation. In contrast, the (12,12) system exhibits a monotonic decrease in $E_\mathrm{interphase}$ from $-1.71\times10^{3}$ to $-2.01\times10^{3}$~kcal~mol$^{-1}$, indicating progressive energy accumulation within a constrained interphase until catastrophic failure near $\varepsilon\!\approx\!0.20$. This opposite slope polarity quantitatively captures the curvature‐controlled direction of energy flow—outward and dissipative in (10,10), inward and confining in (12,12).

Mechanistically, the origin of this polarity is geometric rather than chemical. On the curved (10,10) nanotube, PDA chains experience azimuthal misalignment and partial $\pi$–$\pi$ torsion, creating a deformable interfacial network that can slip, twist, and redistribute stress through frictional damping and transient electrostatic reorientation. On the flatter (12,12) surface, the same PDA monomers assemble in a more ordered $\pi$–$\pi$ configuration, forming a rigid, cooperative adhesive layer with minimal internal mobility. The former thus behaves as an energy‐dissipative reservoir, while the latter functions as an energy‐accumulating barrier. In essence, curvature dictates how PDA mediates mechanical work: high curvature introduces spatially discrete adsorption sites that decouple load through localized vdW dissipation, whereas low curvature enforces collective load bearing through continuous $\pi$–$\pi$ coupling. Therefore, the root cause of the curvature–functionalization puzzle lies in the curvature‐induced topological transition in PDA adsorption, which transforms the interphase from a frictional, energy‐releasing network to a cohesive, energy‐storing one.

\subsection{Integrated Workflow for Curvature–Functionalization Analysis}

Figure~\ref{fig:workflow} summarizes the full workflow adopted in this study, linking model construction, reactive simulations, structural–mechanical analysis, and energetic decomposition into a unified curvature–functionalization evaluation framework.

\medskip
\textbf{Stage 1: CNT–centered polymer packing.}
Atomistic models of PVA–CNT nanocomposites were constructed by embedding (10,10) and (12,12) armchair CNTs within a PVA matrix containing pristine, HNO$_3$-treated, or PDA-functionalized surfaces. Packmol and custom scripts were used to randomize chain placement, enforce realistic density targets, and maintain curvature-consistent local packing around each CNT.

\medskip
\textbf{Stage 2: Reactive molecular dynamics simulation.}
Each configuration underwent energy minimization and equilibration under ReaxFF, followed by uniaxial tensile loading along the CNT axis. These simulations captured bond rearrangement, charge redistribution, interfacial load transfer, and nanoscale deformation modes. Stress–strain responses, potential-energy trajectories, mean-squared displacement (MSD), and porosity evolution were extracted as primary structural, mechanical, and transport descriptors.

\medskip
\textbf{Stage 3: Emergence of curvature-dependent anomalies.}
The mechanical, transport, and structural quantities obtained from the simulations
(modulus, toughness, porosity, and chain mobility) were examined through direct
cross-comparisons across all six CNT–PVA systems. This qualitative analysis revealed
a pronounced curvature-dependent divergence: PDA functionalization produced 
simultaneous increases in porosity, mobility, and toughness for the high-curvature 
(10,10) composite, while the low-curvature (12,12) counterpart remained comparatively 
rigid and weakly responsive to the same treatment. This opposing behavior—arising 
despite identical chemistry and loading conditions—constituted the central puzzle 
motivating the subsequent energetic analysis.

\medskip
\textbf{Stage 4: Energetic decomposition and puzzle resolution.}
To uncover the mechanistic origin of this divergence, the total potential energy was decomposed into bonded, van der Waals, electrostatic, and interphase interaction components. This analysis demonstrated that curvature governs the direction of intercomponent energy flow—dissipative and outward in the (10,10) system, confining and inward in the (12,12) system—thereby establishing the concept of a curvature-adaptive interphase.

\medskip
\textbf{Summary.}
The combined packing–simulation–analysis–decomposition workflow provides a coherent framework for isolating how geometry and chemistry jointly regulate interfacial energetics. This integrated approach enabled the resolution of the curvature–functionalization puzzle and established curvature as a decisive design parameter for CNT–polymer interfaces.

\begin{figure*}[htbp]
    \centering
    \includegraphics[width=0.95\textwidth]{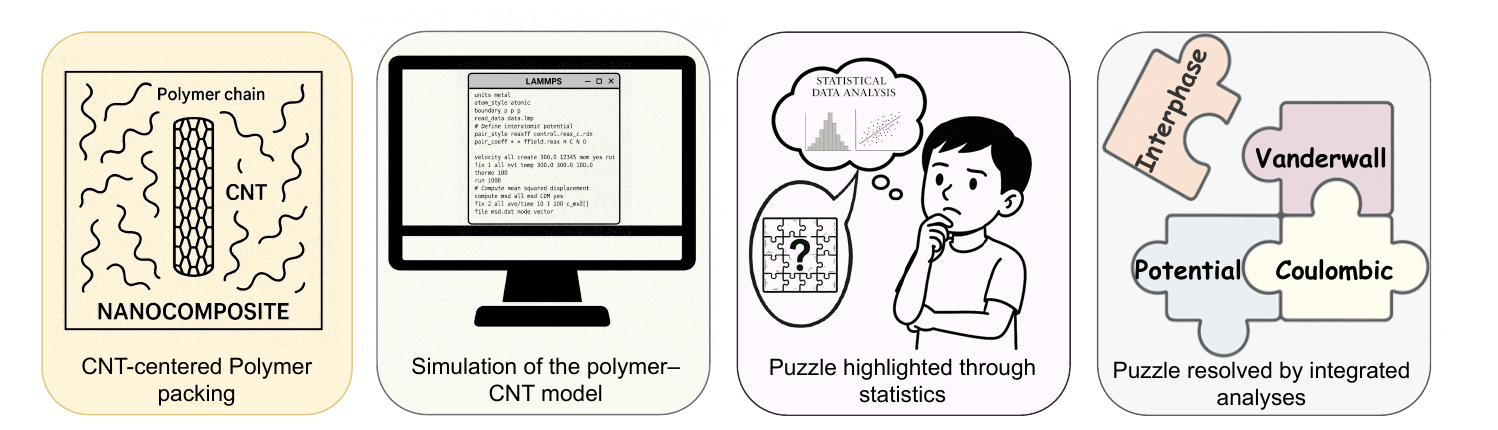}

    \caption{
        \textbf{Overall workflow of the CNT–polymer nanocomposite study.}
        (a) CNT-centered polymer packing. 
        (b) Reactive molecular dynamics simulation using LAMMPS. 
        (c) Data analytics revealing curvature–functionalization interactions. 
        (d) Resolution of the puzzle through interphase energy decomposition combining 
        potential, vdW, electrostatic, and interphase energies.
    }
    \label{fig:workflow}
\end{figure*}

\section*{Conclusion}

This study demonstrates that nanotube curvature governs the polarity of interfacial energy exchange in PDA-functionalized polymer nanocomposites, reversing conventional reinforcement trends between otherwise identical systems. By quantitatively resolving potential, van der Waals, and electrostatic energy evolution during deformation, we reveal that high-curvature \((10,10)\) CNTs dissipate energy through progressive interphase relaxation, whereas low-curvature \((12,12)\) CNTs accumulate it elastically until failure. This energetic inversion explains the apparent contradiction in functionalized CNT--polymer performance and establishes curvature-induced modulation of $\pi$--$\pi$ stacking and interphase energetics as the governing mechanism behind the puzzle. The implications of this work extend well beyond the specific systems studied. By systematically contrasting two representative CNT chiralities and three surface treatments---pristine, HNO$_3$, and PDA---this study demonstrates that curvature can fundamentally redirect interfacial energy flow even under identical chemical functionalization. Although the analysis is limited to a narrow set of geometries and functional groups, the mechanistic origin uncovered here---curvature-induced modulation of $\pi$--$\pi$ stacking geometry and interphase energetics---captures a transferable physical principle that can be generalized to other nanostructured composites. This insight clarifies long-standing inconsistencies in CNT functionalization studies and reframes curvature as a co-equal design variable alongside chemistry. By establishing an atomistic foundation that links surface geometry, interphase energy evolution, and macroscopic mechanical polarity, the present work offers a mechanistic guide for designing next-generation nanocomposites, flexible electronics, and adaptive coatings where controlled energy dissipation and interfacial integrity are critical. Future work should extend this curvature--functionalization framework to a broader range of CNT chiralities, polymer chemistries, and interfacial environments to validate the generality of the observed polarity transition and bridge simulation predictions with experimental measurements.

\bibliography{reference}

\providecommand{\latin}[1]{#1}
\makeatletter
\providecommand{\doi}
  {\begingroup\let\do\@makeother\dospecials
  \catcode`\{=1 \catcode`\}=2 \doi@aux}
\providecommand{\doi@aux}[1]{\endgroup\texttt{#1}}
\makeatother
\providecommand*\mcitethebibliography{\thebibliography}
\csname @ifundefined\endcsname{endmcitethebibliography}  {\let\endmcitethebibliography\endthebibliography}{}
\begin{mcitethebibliography}{66}
\providecommand*\natexlab[1]{#1}
\providecommand*\mciteSetBstSublistMode[1]{}
\providecommand*\mciteSetBstMaxWidthForm[2]{}
\providecommand*\mciteBstWouldAddEndPuncttrue
  {\def\EndOfBibitem{\unskip.}}
\providecommand*\mciteBstWouldAddEndPunctfalse
  {\let\EndOfBibitem\relax}
\providecommand*\mciteSetBstMidEndSepPunct[3]{}
\providecommand*\mciteSetBstSublistLabelBeginEnd[3]{}
\providecommand*\EndOfBibitem{}
\mciteSetBstSublistMode{f}
\mciteSetBstMaxWidthForm{subitem}{(\alph{mcitesubitemcount})}
\mciteSetBstSublistLabelBeginEnd
  {\mcitemaxwidthsubitemform\space}
  {\relax}
  {\relax}

\bibitem[Imtiaz \latin{et~al.}(2018)Imtiaz, Siddiq, Kausar, Muntha, Ambreen, and Bibi]{imtiaz2018review}
Imtiaz,~S.; Siddiq,~M.; Kausar,~A.; Muntha,~S.~T.; Ambreen,~J.; Bibi,~I. A review featuring fabrication, properties and applications of carbon nanotubes (CNTs) reinforced polymer and epoxy nanocomposites. \emph{Chinese Journal of Polymer Science} \textbf{2018}, \emph{36}, 445--461\relax
\mciteBstWouldAddEndPuncttrue
\mciteSetBstMidEndSepPunct{\mcitedefaultmidpunct}
{\mcitedefaultendpunct}{\mcitedefaultseppunct}\relax
\EndOfBibitem
\bibitem[Miyagawa \latin{et~al.}(2005)Miyagawa, Misra, and Mohanty]{miyagawa2005mechanical}
Miyagawa,~H.; Misra,~M.; Mohanty,~A.~K. Mechanical properties of carbon nanotubes and their polymer nanocomposites. \emph{Journal of Nanoscience and Nanotechnology} \textbf{2005}, \emph{5}, 1593--1615\relax
\mciteBstWouldAddEndPuncttrue
\mciteSetBstMidEndSepPunct{\mcitedefaultmidpunct}
{\mcitedefaultendpunct}{\mcitedefaultseppunct}\relax
\EndOfBibitem
\bibitem[Kausar(2021)]{kausar2021holistic}
Kausar,~A. Holistic analysis of nanocomposites of carbon nanotube with polypropylene. \emph{Materials Research Innovations} \textbf{2021}, \emph{25}, 186--197\relax
\mciteBstWouldAddEndPuncttrue
\mciteSetBstMidEndSepPunct{\mcitedefaultmidpunct}
{\mcitedefaultendpunct}{\mcitedefaultseppunct}\relax
\EndOfBibitem
\bibitem[Hasan \latin{et~al.}(2024)Hasan, Islam, and Masud]{hasan2024tailoring}
Hasan,~M.; Islam,~K.; Masud,~A. Tailoring Polyamide Nanocomposites: The Synergistic Effects of SWCNT Chirality and Maleic Anhydride Grafting. \emph{ACS Applied Engineering Materials} \textbf{2024}, \emph{2}, 1593--1602\relax
\mciteBstWouldAddEndPuncttrue
\mciteSetBstMidEndSepPunct{\mcitedefaultmidpunct}
{\mcitedefaultendpunct}{\mcitedefaultseppunct}\relax
\EndOfBibitem
\bibitem[Zheng \latin{et~al.}(2018)Zheng, Wang, Dong, Wu, and Zhang]{zheng2018high}
Zheng,~Y.; Wang,~R.; Dong,~X.; Wu,~L.; Zhang,~X. High strength conductive polyamide 6 nanocomposites reinforced by prebuilt three-dimensional carbon nanotube networks. \emph{ACS Applied Materials \& Interfaces} \textbf{2018}, \emph{10}, 28103--28111\relax
\mciteBstWouldAddEndPuncttrue
\mciteSetBstMidEndSepPunct{\mcitedefaultmidpunct}
{\mcitedefaultendpunct}{\mcitedefaultseppunct}\relax
\EndOfBibitem
\bibitem[Zare and Rhee(2021)Zare, and Rhee]{zare2021effects}
Zare,~Y.; Rhee,~K.~Y. Effects of interfacial shear strength on the operative aspects of interphase section and tensile strength of carbon-nanotube-filled system: A modeling study. \emph{Results in Physics} \textbf{2021}, \emph{26}, 104428\relax
\mciteBstWouldAddEndPuncttrue
\mciteSetBstMidEndSepPunct{\mcitedefaultmidpunct}
{\mcitedefaultendpunct}{\mcitedefaultseppunct}\relax
\EndOfBibitem
\bibitem[Mohd~Nurazzi \latin{et~al.}(2021)Mohd~Nurazzi, Asyraf, Khalina, Abdullah, Sabaruddin, Kamarudin, Ahmad, Mahat, Lee, Aisyah, \latin{et~al.} others]{mohd2021fabrication}
Mohd~Nurazzi,~N.; Asyraf,~M.~M.; Khalina,~A.; Abdullah,~N.; Sabaruddin,~F.~A.; Kamarudin,~S.~H.; Ahmad,~S.; Mahat,~A.~M.; Lee,~C.~L.; Aisyah,~H.; others Fabrication, functionalization, and application of carbon nanotube-reinforced polymer composite: An overview. \emph{Polymers} \textbf{2021}, \emph{13}, 1047\relax
\mciteBstWouldAddEndPuncttrue
\mciteSetBstMidEndSepPunct{\mcitedefaultmidpunct}
{\mcitedefaultendpunct}{\mcitedefaultseppunct}\relax
\EndOfBibitem
\bibitem[Falara \latin{et~al.}(2022)Falara, Zourou, and Kordatos]{falara2022recent}
Falara,~P.~P.; Zourou,~A.; Kordatos,~K.~V. Recent advances in Carbon Dots/2-D hybrid materials. \emph{Carbon} \textbf{2022}, \emph{195}, 219--245\relax
\mciteBstWouldAddEndPuncttrue
\mciteSetBstMidEndSepPunct{\mcitedefaultmidpunct}
{\mcitedefaultendpunct}{\mcitedefaultseppunct}\relax
\EndOfBibitem
\bibitem[Sandhu(2007)]{sandhu2007career}
Sandhu,~A. A career in carbon. \emph{Nature Nanotechnology} \textbf{2007}, \emph{2}, 590--591\relax
\mciteBstWouldAddEndPuncttrue
\mciteSetBstMidEndSepPunct{\mcitedefaultmidpunct}
{\mcitedefaultendpunct}{\mcitedefaultseppunct}\relax
\EndOfBibitem
\bibitem[Behabtu \latin{et~al.}(2013)Behabtu, Young, Tsentalovich, Kleinerman, Wang, Ma, Bengio, Ter~Waarbeek, De~Jong, Hoogerwerf, \latin{et~al.} others]{behabtu2013strong}
Behabtu,~N.; Young,~C.~C.; Tsentalovich,~D.~E.; Kleinerman,~O.; Wang,~X.; Ma,~A.~W.; Bengio,~E.~A.; Ter~Waarbeek,~R.~F.; De~Jong,~J.~J.; Hoogerwerf,~R.~E.; others Strong, light, multifunctional fibers of carbon nanotubes with ultrahigh conductivity. \emph{science} \textbf{2013}, \emph{339}, 182--186\relax
\mciteBstWouldAddEndPuncttrue
\mciteSetBstMidEndSepPunct{\mcitedefaultmidpunct}
{\mcitedefaultendpunct}{\mcitedefaultseppunct}\relax
\EndOfBibitem
\bibitem[Zhou \latin{et~al.}(2020)Zhou, Zhong, Zhao, Meng, and Qi]{zhou2020role}
Zhou,~J.; Zhong,~K.; Zhao,~C.; Meng,~H.; Qi,~L. On the role of carbon nanotubes addition in carbon fiber-reinforced magnesium matrix composites. \emph{Journal of Materials Science} \textbf{2020}, \emph{55}, 16940--16953\relax
\mciteBstWouldAddEndPuncttrue
\mciteSetBstMidEndSepPunct{\mcitedefaultmidpunct}
{\mcitedefaultendpunct}{\mcitedefaultseppunct}\relax
\EndOfBibitem
\bibitem[Lekshmi \latin{et~al.}(2020)Lekshmi, Sana, Nguyen, Nguyen, Nguyen, Le, and Peng]{lekshmi2020recent}
Lekshmi,~G.; Sana,~S.~S.; Nguyen,~V.-H.; Nguyen,~T. H.~C.; Nguyen,~C.~C.; Le,~Q.~V.; Peng,~W. Recent progress in carbon nanotube polymer composites in tissue engineering and regeneration. \emph{International Journal of Molecular Sciences} \textbf{2020}, \emph{21}, 6440\relax
\mciteBstWouldAddEndPuncttrue
\mciteSetBstMidEndSepPunct{\mcitedefaultmidpunct}
{\mcitedefaultendpunct}{\mcitedefaultseppunct}\relax
\EndOfBibitem
\bibitem[Yang \latin{et~al.}(2015)Yang, Xu, Liu, Liu, Li, Zhang, and Lv]{yang2015fabrication}
Yang,~Z.; Xu,~D.; Liu,~J.; Liu,~J.; Li,~L.; Zhang,~L.; Lv,~J. Fabrication and characterization of poly (vinyl alcohol)/carbon nanotube melt-spinning composites fiber. \emph{Progress in Natural Science: Materials International} \textbf{2015}, \emph{25}, 437--444\relax
\mciteBstWouldAddEndPuncttrue
\mciteSetBstMidEndSepPunct{\mcitedefaultmidpunct}
{\mcitedefaultendpunct}{\mcitedefaultseppunct}\relax
\EndOfBibitem
\bibitem[Yim \latin{et~al.}(2025)Yim, Yoon, Kim, Lee, Chung, and Kim]{yim2025carbon}
Yim,~Y.-J.; Yoon,~Y.-H.; Kim,~S.-H.; Lee,~J.-H.; Chung,~D.-C.; Kim,~B.-J. Carbon nanotube/polymer composites for functional applications. \emph{Polymers} \textbf{2025}, \emph{17}, 119\relax
\mciteBstWouldAddEndPuncttrue
\mciteSetBstMidEndSepPunct{\mcitedefaultmidpunct}
{\mcitedefaultendpunct}{\mcitedefaultseppunct}\relax
\EndOfBibitem
\bibitem[Li \latin{et~al.}(2022)Li, Bai, Luo, Ding, Li, and Liang]{li2022cnt}
Li,~S.; Bai,~L.; Luo,~X.; Ding,~J.; Li,~G.; Liang,~H. A CNT/PVA film supported TFC membranes for improvement of mechanical properties and chemical cleaning stability: A new insight to an alternative to the polymeric support. \emph{Journal of Membrane Science} \textbf{2022}, \emph{658}, 120753\relax
\mciteBstWouldAddEndPuncttrue
\mciteSetBstMidEndSepPunct{\mcitedefaultmidpunct}
{\mcitedefaultendpunct}{\mcitedefaultseppunct}\relax
\EndOfBibitem
\bibitem[Liu \latin{et~al.}(2023)Liu, Li, Geng, Cao, Tian, Li, Bin, Qian, and Geng]{liu2023modified}
Liu,~X.-L.; Li,~M.; Geng,~W.-H.; Cao,~W.; Tian,~Y.-H.; Li,~T.-Y.; Bin,~P.-S.; Qian,~P.-F.; Geng,~H.-Z. Modified carbon nanotubes/polyvinyl alcohol composite electrothermal films. \emph{Surfaces and Interfaces} \textbf{2023}, \emph{36}, 102540\relax
\mciteBstWouldAddEndPuncttrue
\mciteSetBstMidEndSepPunct{\mcitedefaultmidpunct}
{\mcitedefaultendpunct}{\mcitedefaultseppunct}\relax
\EndOfBibitem
\bibitem[Yang \latin{et~al.}(2011)Yang, Xie, and Mai]{yang2011functionalization}
Yang,~Y.-K.; Xie,~X.-L.; Mai,~Y.-W. \emph{Polymer--Carbon Nanotube Composites}; Elsevier, 2011; pp 55--91\relax
\mciteBstWouldAddEndPuncttrue
\mciteSetBstMidEndSepPunct{\mcitedefaultmidpunct}
{\mcitedefaultendpunct}{\mcitedefaultseppunct}\relax
\EndOfBibitem
\bibitem[Siwal \latin{et~al.}(2020)Siwal, Zhang, Devi, and Thakur]{siwal2020carbon}
Siwal,~S.~S.; Zhang,~Q.; Devi,~N.; Thakur,~V.~K. Carbon-based polymer nanocomposite for high-performance energy storage applications. \emph{Polymers} \textbf{2020}, \emph{12}, 505\relax
\mciteBstWouldAddEndPuncttrue
\mciteSetBstMidEndSepPunct{\mcitedefaultmidpunct}
{\mcitedefaultendpunct}{\mcitedefaultseppunct}\relax
\EndOfBibitem
\bibitem[Sabet(2024)]{sabet2024innovative}
Sabet,~M. Innovative synthesis and functionalization of carbon nanotube-polymer composites: Enhanced performance, mechanisms, and emerging applications. \emph{Materials Science and Technology} \textbf{2024}, 02670836251356021\relax
\mciteBstWouldAddEndPuncttrue
\mciteSetBstMidEndSepPunct{\mcitedefaultmidpunct}
{\mcitedefaultendpunct}{\mcitedefaultseppunct}\relax
\EndOfBibitem
\bibitem[Suslova \latin{et~al.}(2019)Suslova, Arkhipova, Kalashnik, Ivanov, Savilov, Xia, and Lunin]{suslova2019effect}
Suslova,~E.; Arkhipova,~E.; Kalashnik,~A.; Ivanov,~A.; Savilov,~S.; Xia,~H.; Lunin,~V. Effect of the functionalization of nitrogen-doped carbon nanotubes on electrical conductivity. \emph{Russian Journal of Physical Chemistry A} \textbf{2019}, \emph{93}, 1952--1956\relax
\mciteBstWouldAddEndPuncttrue
\mciteSetBstMidEndSepPunct{\mcitedefaultmidpunct}
{\mcitedefaultendpunct}{\mcitedefaultseppunct}\relax
\EndOfBibitem
\bibitem[Zeng \latin{et~al.}(2025)Zeng, Wu, Li, Zhou, Zheng, Feng, Yang, Ren, Lu, Ying, \latin{et~al.} others]{zeng2025artificial}
Zeng,~M.; Wu,~Y.; Li,~N.; Zhou,~J.; Zheng,~X.; Feng,~R.; Yang,~Z.; Ren,~X.; Lu,~S.; Ying,~J.; others Artificial nacre based on polydopamine functionalized graphene oxide nanosheets constrained palladium nanocluster with enhanced mechanical properties and catalytical functionalities. \emph{International Journal of Biological Macromolecules} \textbf{2025}, 147536\relax
\mciteBstWouldAddEndPuncttrue
\mciteSetBstMidEndSepPunct{\mcitedefaultmidpunct}
{\mcitedefaultendpunct}{\mcitedefaultseppunct}\relax
\EndOfBibitem
\bibitem[Sharma and Sharma(2025)Sharma, and Sharma]{sharma2025enhancing}
Sharma,~P.; Sharma,~S.~P. Enhancing interfacial adhesion in carbon fiber-reinforced composites through polydopamine-assisted nanomaterial coatings. \emph{Polymer Composites} \textbf{2025}, \emph{46}, 6768--6784\relax
\mciteBstWouldAddEndPuncttrue
\mciteSetBstMidEndSepPunct{\mcitedefaultmidpunct}
{\mcitedefaultendpunct}{\mcitedefaultseppunct}\relax
\EndOfBibitem
\bibitem[Wu \latin{et~al.}(2025)Wu, Jin, Ma, Xiao, Li, Zhang, and Zhu]{wu2025two}
Wu,~Q.; Jin,~D.; Ma,~Q.; Xiao,~B.; Li,~Y.; Zhang,~Y.; Zhu,~J. Two geometries of polydopamine synergistically enhance carbon fiber-epoxy interfacial adhesion. \emph{Composites Part B: Engineering} \textbf{2025}, 112768\relax
\mciteBstWouldAddEndPuncttrue
\mciteSetBstMidEndSepPunct{\mcitedefaultmidpunct}
{\mcitedefaultendpunct}{\mcitedefaultseppunct}\relax
\EndOfBibitem
\bibitem[Jiang \latin{et~al.}(2025)Jiang, Sadia, Yu, Zheng, and Dong]{jiang2025encapsulating}
Jiang,~W.-J.; Sadia,~H.; Yu,~D.-F.; Zheng,~P.-F.; Dong,~Z.-H. Encapsulating PDA-wrapped carbon fibers in waterborne epoxy zinc-rich coatings to enhance the utilization and bonding strength of Zn particles. \emph{Colloids and Surfaces A: Physicochemical and Engineering Aspects} \textbf{2025}, \emph{722}, 137143\relax
\mciteBstWouldAddEndPuncttrue
\mciteSetBstMidEndSepPunct{\mcitedefaultmidpunct}
{\mcitedefaultendpunct}{\mcitedefaultseppunct}\relax
\EndOfBibitem
\bibitem[Frenzel \latin{et~al.}(2025)Frenzel, Drechsler, Zimmerer, Synytska, Rezaie, Ahmed, Liebscher, and Mechtcherine]{frenzel2025combined}
Frenzel,~R.; Drechsler,~A.; Zimmerer,~C.; Synytska,~A.; Rezaie,~A.~B.; Ahmed,~A.~H.; Liebscher,~M.; Mechtcherine,~V. Combined polydopamine/polyelectrolyte modification of polyethylene fibers to promote adhesion in fiber-reinforced cement composites. \emph{Colloids and Surfaces A: Physicochemical and Engineering Aspects} \textbf{2025}, 138175\relax
\mciteBstWouldAddEndPuncttrue
\mciteSetBstMidEndSepPunct{\mcitedefaultmidpunct}
{\mcitedefaultendpunct}{\mcitedefaultseppunct}\relax
\EndOfBibitem
\bibitem[Cai \latin{et~al.}(2018)Cai, Hou, Jiang, and Dong]{cai2018polydopamine}
Cai,~G.; Hou,~J.; Jiang,~D.; Dong,~Z. Polydopamine-wrapped carbon nanotubes to improve the corrosion barrier of polyurethane coating. \emph{RSC advances} \textbf{2018}, \emph{8}, 23727--23741\relax
\mciteBstWouldAddEndPuncttrue
\mciteSetBstMidEndSepPunct{\mcitedefaultmidpunct}
{\mcitedefaultendpunct}{\mcitedefaultseppunct}\relax
\EndOfBibitem
\bibitem[Demirci \latin{et~al.}(2021)Demirci, Sahiner, Suner, and Sahiner]{demirci2021improved}
Demirci,~S.; Sahiner,~M.; Suner,~S.~S.; Sahiner,~N. Improved biomedical properties of polydopamine-coated carbon nanotubes. \emph{Micromachines} \textbf{2021}, \emph{12}, 1280\relax
\mciteBstWouldAddEndPuncttrue
\mciteSetBstMidEndSepPunct{\mcitedefaultmidpunct}
{\mcitedefaultendpunct}{\mcitedefaultseppunct}\relax
\EndOfBibitem
\bibitem[Kim and Kim(2020)Kim, and Kim]{kim2020adsorption}
Kim,~H.; Kim,~G. Adsorption properties of dopamine derivatives using carbon nanotubes: A first-principles study. \emph{Applied Surface Science} \textbf{2020}, \emph{501}, 144249\relax
\mciteBstWouldAddEndPuncttrue
\mciteSetBstMidEndSepPunct{\mcitedefaultmidpunct}
{\mcitedefaultendpunct}{\mcitedefaultseppunct}\relax
\EndOfBibitem
\bibitem[Thompson \latin{et~al.}(2022)Thompson, Aktulga, Berger, Bolintineanu, Brown, Crozier, {in 't Veld}, Kohlmeyer, Moore, Nguyen, Shan, Stevens, Tranchida, Trott, and Plimpton]{THOMPSON2022108171}
Thompson,~A.~P.; Aktulga,~H.~M.; Berger,~R.; Bolintineanu,~D.~S.; Brown,~W.~M.; Crozier,~P.~S.; {in 't Veld},~P.~J.; Kohlmeyer,~A.; Moore,~S.~G.; Nguyen,~T.~D.; Shan,~R.; Stevens,~M.~J.; Tranchida,~J.; Trott,~C.; Plimpton,~S.~J. LAMMPS - a flexible simulation tool for particle-based materials modeling at the atomic, meso, and continuum scales. \emph{Computer Physics Communications} \textbf{2022}, \emph{271}, 108171\relax
\mciteBstWouldAddEndPuncttrue
\mciteSetBstMidEndSepPunct{\mcitedefaultmidpunct}
{\mcitedefaultendpunct}{\mcitedefaultseppunct}\relax
\EndOfBibitem
\bibitem[Van~Duin \latin{et~al.}(2001)Van~Duin, Dasgupta, Lorant, and Goddard]{van2001reaxff}
Van~Duin,~A.~C.; Dasgupta,~S.; Lorant,~F.; Goddard,~W.~A. ReaxFF: a reactive force field for hydrocarbons. \emph{The Journal of Physical Chemistry A} \textbf{2001}, \emph{105}, 9396--9409\relax
\mciteBstWouldAddEndPuncttrue
\mciteSetBstMidEndSepPunct{\mcitedefaultmidpunct}
{\mcitedefaultendpunct}{\mcitedefaultseppunct}\relax
\EndOfBibitem
\bibitem[Chenoweth \latin{et~al.}(2008)Chenoweth, Van~Duin, and Goddard]{chenoweth2008reaxff}
Chenoweth,~K.; Van~Duin,~A.~C.; Goddard,~W.~A. ReaxFF reactive force field for molecular dynamics simulations of hydrocarbon oxidation. \emph{The Journal of Physical Chemistry A} \textbf{2008}, \emph{112}, 1040--1053\relax
\mciteBstWouldAddEndPuncttrue
\mciteSetBstMidEndSepPunct{\mcitedefaultmidpunct}
{\mcitedefaultendpunct}{\mcitedefaultseppunct}\relax
\EndOfBibitem
\bibitem[Sharma \latin{et~al.}(2012)Sharma, Saxena, and Shukla]{sharma2012effect}
Sharma,~K.; Saxena,~K.~K.; Shukla,~M. Effect of multiple Stone-Wales and Vacancy defects on the mechanical behavior of carbon nanotubes using Molecular Dynamics. \emph{Procedia Engineering} \textbf{2012}, \emph{38}, 3373--3380\relax
\mciteBstWouldAddEndPuncttrue
\mciteSetBstMidEndSepPunct{\mcitedefaultmidpunct}
{\mcitedefaultendpunct}{\mcitedefaultseppunct}\relax
\EndOfBibitem
\bibitem[Rissanou \latin{et~al.}(2018)Rissanou, Ba{\v{c}}ov{\'a}, Power, and Harmandaris]{rissanou2018atomistic}
Rissanou,~N.; Ba{\v{c}}ov{\'a},~P.; Power,~A.; Harmandaris,~V. Atomistic molecular dynamics simulations of polymer/graphene nanostructured systems. \emph{Materials Today: Proceedings} \textbf{2018}, \emph{5}, 27472--27481\relax
\mciteBstWouldAddEndPuncttrue
\mciteSetBstMidEndSepPunct{\mcitedefaultmidpunct}
{\mcitedefaultendpunct}{\mcitedefaultseppunct}\relax
\EndOfBibitem
\bibitem[Hashim \latin{et~al.}(2014)Hashim, El-Mekawey, El-Kashef, and Ghazy]{hashim2014determination}
Hashim,~H.; El-Mekawey,~F.; El-Kashef,~H.; Ghazy,~R. Determination of scattering parameters of polyvinyl alcohol by static laser scattering. \emph{Beni-suef university journal of basic and applied sciences} \textbf{2014}, \emph{3}, 203--208\relax
\mciteBstWouldAddEndPuncttrue
\mciteSetBstMidEndSepPunct{\mcitedefaultmidpunct}
{\mcitedefaultendpunct}{\mcitedefaultseppunct}\relax
\EndOfBibitem
\bibitem[Mori \latin{et~al.}(2005)Mori, Hirai, Ogata, Akita, and Nakayama]{mori2005chirality}
Mori,~H.; Hirai,~Y.; Ogata,~S.; Akita,~S.; Nakayama,~Y. Chirality dependence of mechanical properties of single-walled carbon nanotubes under axial tensile strain. \emph{Japanese Journal of Applied Physics} \textbf{2005}, \emph{44}, L1307\relax
\mciteBstWouldAddEndPuncttrue
\mciteSetBstMidEndSepPunct{\mcitedefaultmidpunct}
{\mcitedefaultendpunct}{\mcitedefaultseppunct}\relax
\EndOfBibitem
\bibitem[Natsuki \latin{et~al.}(2004)Natsuki, Tantrakarn, and Endo]{natsuki2004effects}
Natsuki,~T.; Tantrakarn,~K.; Endo,~M. Effects of carbon nanotube structures on mechanical properties. \emph{Applied Physics A} \textbf{2004}, \emph{79}, 117--124\relax
\mciteBstWouldAddEndPuncttrue
\mciteSetBstMidEndSepPunct{\mcitedefaultmidpunct}
{\mcitedefaultendpunct}{\mcitedefaultseppunct}\relax
\EndOfBibitem
\bibitem[Ketolainen \latin{et~al.}(2018)Ketolainen, Havu, J{\'o}nsson, and Puska]{ketolainen2018electronic}
Ketolainen,~T.; Havu,~V.; J{\'o}nsson,~E.; Puska,~M. Electronic transport properties of carbon-nanotube networks: The effect of nitrate doping on intratube and intertube conductances. \emph{Physical Review Applied} \textbf{2018}, \emph{9}, 034010\relax
\mciteBstWouldAddEndPuncttrue
\mciteSetBstMidEndSepPunct{\mcitedefaultmidpunct}
{\mcitedefaultendpunct}{\mcitedefaultseppunct}\relax
\EndOfBibitem
\bibitem[Wang and Chen(2019)Wang, and Chen]{wang2019preparation}
Wang,~Y.; Chen,~J. Preparation and characterization of polydopamine-modified Ni/carbon nanotubes friction composite coating. \emph{Coatings} \textbf{2019}, \emph{9}, 596\relax
\mciteBstWouldAddEndPuncttrue
\mciteSetBstMidEndSepPunct{\mcitedefaultmidpunct}
{\mcitedefaultendpunct}{\mcitedefaultseppunct}\relax
\EndOfBibitem
\bibitem[Plimpton(1995)]{plimpton1995fast}
Plimpton,~S. Fast parallel algorithms for short-range molecular dynamics. \emph{Journal of computational physics} \textbf{1995}, \emph{117}, 1--19\relax
\mciteBstWouldAddEndPuncttrue
\mciteSetBstMidEndSepPunct{\mcitedefaultmidpunct}
{\mcitedefaultendpunct}{\mcitedefaultseppunct}\relax
\EndOfBibitem
\bibitem[Dupont and Germann(2012)Dupont, and Germann]{dupont2012strain}
Dupont,~V.; Germann,~T.~C. Strain rate and orientation dependencies of the strength of single crystalline copper under compression. \emph{Physical Review B—Condensed Matter and Materials Physics} \textbf{2012}, \emph{86}, 134111\relax
\mciteBstWouldAddEndPuncttrue
\mciteSetBstMidEndSepPunct{\mcitedefaultmidpunct}
{\mcitedefaultendpunct}{\mcitedefaultseppunct}\relax
\EndOfBibitem
\bibitem[Youssef \latin{et~al.}(2024)Youssef, Reda, and Harmandaris]{youssef2024unraveling}
Youssef,~A.~A.; Reda,~H.; Harmandaris,~V. Unraveling the Effect of Strain Rate and Temperature on the Heterogeneous Mechanical Behavior of Polymer Nanocomposites via Atomistic Simulations and Continuum Models. \emph{Polymers} \textbf{2024}, \emph{16}, 2530\relax
\mciteBstWouldAddEndPuncttrue
\mciteSetBstMidEndSepPunct{\mcitedefaultmidpunct}
{\mcitedefaultendpunct}{\mcitedefaultseppunct}\relax
\EndOfBibitem
\bibitem[Huang \latin{et~al.}(2022)Huang, Zhou, and Liu]{huang2022interphase}
Huang,~J.; Zhou,~J.; Liu,~M. Interphase in polymer nanocomposites. \emph{JACS Au} \textbf{2022}, \emph{2}, 280--291\relax
\mciteBstWouldAddEndPuncttrue
\mciteSetBstMidEndSepPunct{\mcitedefaultmidpunct}
{\mcitedefaultendpunct}{\mcitedefaultseppunct}\relax
\EndOfBibitem
\bibitem[Kumar(2023)]{kumar2023characterizing}
Kumar,~S. Characterizing interphase region in CNT/nylon-6 composites using molecular dynamics simulation. \emph{Bulletin of Materials Science} \textbf{2023}, \emph{46}, 24\relax
\mciteBstWouldAddEndPuncttrue
\mciteSetBstMidEndSepPunct{\mcitedefaultmidpunct}
{\mcitedefaultendpunct}{\mcitedefaultseppunct}\relax
\EndOfBibitem
\bibitem[Su \latin{et~al.}(2020)Su, Chen, Ju, Chen, Shih, Pan, and You]{su2020mechanical}
Su,~C.-H.; Chen,~H.-L.; Ju,~S.-P.; Chen,~H.-Y.; Shih,~C.-W.; Pan,~C.-T.; You,~T.-D. The mechanical behaviors of polyethylene/silver nanoparticle composites: an insight from molecular dynamics study. \emph{Scientific Reports} \textbf{2020}, \emph{10}, 7600\relax
\mciteBstWouldAddEndPuncttrue
\mciteSetBstMidEndSepPunct{\mcitedefaultmidpunct}
{\mcitedefaultendpunct}{\mcitedefaultseppunct}\relax
\EndOfBibitem
\bibitem[Salvetat \latin{et~al.}(1999)Salvetat, Briggs, Bonard, Bacsa, Kulik, St{\"o}ckli, Burnham, and Forr{\'o}]{salvetat1999elastic}
Salvetat,~J.-P.; Briggs,~G. A.~D.; Bonard,~J.-M.; Bacsa,~R.~R.; Kulik,~A.~J.; St{\"o}ckli,~T.; Burnham,~N.~A.; Forr{\'o},~L. Elastic and shear moduli of single-walled carbon nanotube ropes. \emph{Physical review letters} \textbf{1999}, \emph{82}, 944\relax
\mciteBstWouldAddEndPuncttrue
\mciteSetBstMidEndSepPunct{\mcitedefaultmidpunct}
{\mcitedefaultendpunct}{\mcitedefaultseppunct}\relax
\EndOfBibitem
\bibitem[Yu \latin{et~al.}(2000)Yu, Lourie, Dyer, Moloni, Kelly, and Ruoff]{yu2000strength}
Yu,~M.-F.; Lourie,~O.; Dyer,~M.~J.; Moloni,~K.; Kelly,~T.~F.; Ruoff,~R.~S. Strength and breaking mechanism of multiwalled carbon nanotubes under tensile load. \emph{Science} \textbf{2000}, \emph{287}, 637--640\relax
\mciteBstWouldAddEndPuncttrue
\mciteSetBstMidEndSepPunct{\mcitedefaultmidpunct}
{\mcitedefaultendpunct}{\mcitedefaultseppunct}\relax
\EndOfBibitem
\bibitem[Sharma \latin{et~al.}(2013)Sharma, Chandra, Kumar, and Kumar]{sharma2013molecular}
Sharma,~S.; Chandra,~R.; Kumar,~P.; Kumar,~N. Molecular dynamics simulation of carbon nanotubes. \emph{Nanoscience and Technology: An International Journal} \textbf{2013}, \emph{4}\relax
\mciteBstWouldAddEndPuncttrue
\mciteSetBstMidEndSepPunct{\mcitedefaultmidpunct}
{\mcitedefaultendpunct}{\mcitedefaultseppunct}\relax
\EndOfBibitem
\bibitem[WenXing \latin{et~al.}(2004)WenXing, ChangChun, and WanZhao]{wenxing2004simulation}
WenXing,~B.; ChangChun,~Z.; WanZhao,~C. Simulation of Young's modulus of single-walled carbon nanotubes by molecular dynamics. \emph{Physica B: Condensed Matter} \textbf{2004}, \emph{352}, 156--163\relax
\mciteBstWouldAddEndPuncttrue
\mciteSetBstMidEndSepPunct{\mcitedefaultmidpunct}
{\mcitedefaultendpunct}{\mcitedefaultseppunct}\relax
\EndOfBibitem
\bibitem[Lu and Bhattacharya(2005)Lu, and Bhattacharya]{lu2005effect}
Lu,~Q.; Bhattacharya,~B. Effect of randomly occurring Stone--Wales defects on mechanical properties of carbonnanotubes using atomistic simulation. \emph{Nanotechnology} \textbf{2005}, \emph{16}, 555\relax
\mciteBstWouldAddEndPuncttrue
\mciteSetBstMidEndSepPunct{\mcitedefaultmidpunct}
{\mcitedefaultendpunct}{\mcitedefaultseppunct}\relax
\EndOfBibitem
\bibitem[Brown(2001)]{brown2001step}
Brown,~A.~M. A step-by-step guide to non-linear regression analysis of experimental data using a Microsoft Excel spreadsheet. \emph{Computer methods and programs in biomedicine} \textbf{2001}, \emph{65}, 191--200\relax
\mciteBstWouldAddEndPuncttrue
\mciteSetBstMidEndSepPunct{\mcitedefaultmidpunct}
{\mcitedefaultendpunct}{\mcitedefaultseppunct}\relax
\EndOfBibitem
\bibitem[Moon \latin{et~al.}(2025)Moon, Mim, Billah, and Masud]{moon2025synthesis}
Moon,~M.; Mim,~S.~R.; Billah,~M.~M.; Masud,~A. Synthesis and characterization of surface modified MWCNTs reinforced PVA composite films. \emph{Heliyon} \textbf{2025}, \emph{11}\relax
\mciteBstWouldAddEndPuncttrue
\mciteSetBstMidEndSepPunct{\mcitedefaultmidpunct}
{\mcitedefaultendpunct}{\mcitedefaultseppunct}\relax
\EndOfBibitem
\bibitem[Wunderle \latin{et~al.}(2009)Wunderle, Dermitzaki, Holck, Bauer, Walter, Shaik, Ratzke, Faupel, Michel, and Reichl]{wunderle2009molecular}
Wunderle,~B.; Dermitzaki,~E.; Holck,~O.; Bauer,~J.; Walter,~H.; Shaik,~Q.; Ratzke,~K.; Faupel,~F.; Michel,~B.; Reichl,~H. Molecular dynamics approach to structure-property correlation in epoxy resins for thermo-mechanical lifetime modeling. 2009 59th Electronic Components and Technology Conference. 2009; pp 1404--1413\relax
\mciteBstWouldAddEndPuncttrue
\mciteSetBstMidEndSepPunct{\mcitedefaultmidpunct}
{\mcitedefaultendpunct}{\mcitedefaultseppunct}\relax
\EndOfBibitem
\bibitem[Boyd and Pant(1991)Boyd, and Pant]{boyd1991molecular}
Boyd,~R.~H.; Pant,~P.~K. Molecular packing and diffusion in polyisobutylene. \emph{Macromolecules} \textbf{1991}, \emph{24}, 6325--6331\relax
\mciteBstWouldAddEndPuncttrue
\mciteSetBstMidEndSepPunct{\mcitedefaultmidpunct}
{\mcitedefaultendpunct}{\mcitedefaultseppunct}\relax
\EndOfBibitem
\bibitem[Chipman(1979)]{chipman1979efficiency}
Chipman,~J.~S. Efficiency of least-squares estimation of linear trend when residuals are autocorrelated. \emph{Econometrica: Journal of the Econometric Society} \textbf{1979}, 115--128\relax
\mciteBstWouldAddEndPuncttrue
\mciteSetBstMidEndSepPunct{\mcitedefaultmidpunct}
{\mcitedefaultendpunct}{\mcitedefaultseppunct}\relax
\EndOfBibitem
\bibitem[Gibson and Ashby(1997)Gibson, and Ashby]{Gibson1997}
Gibson,~L.~J.; Ashby,~M.~F. \emph{Cellular Solids: Structure and Properties}, 2nd ed.; Cambridge University Press, 1997\relax
\mciteBstWouldAddEndPuncttrue
\mciteSetBstMidEndSepPunct{\mcitedefaultmidpunct}
{\mcitedefaultendpunct}{\mcitedefaultseppunct}\relax
\EndOfBibitem
\bibitem[Rice(1968)]{Rice1968}
Rice,~J.~R. A Path Independent Integral and the Approximate Analysis of Strain Concentration by Notches and Cracks. \emph{J. Appl. Mech.} \textbf{1968}, \emph{35}, 379--386\relax
\mciteBstWouldAddEndPuncttrue
\mciteSetBstMidEndSepPunct{\mcitedefaultmidpunct}
{\mcitedefaultendpunct}{\mcitedefaultseppunct}\relax
\EndOfBibitem
\bibitem[Huang \latin{et~al.}(2020)Huang, \latin{et~al.} others]{Huang2020}
Huang,~Y.; others Structure--Property Relationships in CNT/Polymer Nanocomposites. \emph{Compos. Sci. Technol.} \textbf{2020}, \emph{197}, 108259\relax
\mciteBstWouldAddEndPuncttrue
\mciteSetBstMidEndSepPunct{\mcitedefaultmidpunct}
{\mcitedefaultendpunct}{\mcitedefaultseppunct}\relax
\EndOfBibitem
\bibitem[Anderson(2017)]{Anderson2017}
Anderson,~T.~L. \emph{Fracture Mechanics: Fundamentals and Applications}, 4th ed.; CRC Press, 2017\relax
\mciteBstWouldAddEndPuncttrue
\mciteSetBstMidEndSepPunct{\mcitedefaultmidpunct}
{\mcitedefaultendpunct}{\mcitedefaultseppunct}\relax
\EndOfBibitem
\bibitem[Budiansky and Hutchinson(1978)Budiansky, and Hutchinson]{Budiansky1978}
Budiansky,~B.; Hutchinson,~J.~W. Matrix Fracture in Fiber-Reinforced Ceramics. \emph{Int. J. Solids Struct.} \textbf{1978}, \emph{14}, 183--202\relax
\mciteBstWouldAddEndPuncttrue
\mciteSetBstMidEndSepPunct{\mcitedefaultmidpunct}
{\mcitedefaultendpunct}{\mcitedefaultseppunct}\relax
\EndOfBibitem
\bibitem[Millington and Quirk(1961)Millington, and Quirk]{Millington1961}
Millington,~R.~J.; Quirk,~J.~P. Permeability of Porous Solids. \emph{Trans. Faraday Soc.} \textbf{1961}, \emph{57}, 1200--1207\relax
\mciteBstWouldAddEndPuncttrue
\mciteSetBstMidEndSepPunct{\mcitedefaultmidpunct}
{\mcitedefaultendpunct}{\mcitedefaultseppunct}\relax
\EndOfBibitem
\bibitem[Shen \latin{et~al.}(2018)Shen, \latin{et~al.} others]{Shen2018}
Shen,~L.; others Design of Porous Polymer Membranes with Enhanced Diffusivity. \emph{J. Membr. Sci.} \textbf{2018}, \emph{563}, 146--158\relax
\mciteBstWouldAddEndPuncttrue
\mciteSetBstMidEndSepPunct{\mcitedefaultmidpunct}
{\mcitedefaultendpunct}{\mcitedefaultseppunct}\relax
\EndOfBibitem
\bibitem[Skountzos \latin{et~al.}(2018)Skountzos, Mermigkis, and Mavrantzas]{skountzos2018molecular}
Skountzos,~E.~N.; Mermigkis,~P.~G.; Mavrantzas,~V.~G. Molecular dynamics study of an atactic poly (methyl methacrylate)--carbon nanotube nanocomposite. \emph{The Journal of Physical Chemistry B} \textbf{2018}, \emph{122}, 9007--9021\relax
\mciteBstWouldAddEndPuncttrue
\mciteSetBstMidEndSepPunct{\mcitedefaultmidpunct}
{\mcitedefaultendpunct}{\mcitedefaultseppunct}\relax
\EndOfBibitem
\bibitem[Bucknall and Paul(2013)Bucknall, and Paul]{Bucknall2013}
Bucknall,~C.~B.; Paul,~D.~R. Notched Impact Behaviour of Polymer Blends: Dependence of Critical Particle Size on Rubber Volume Fraction. \emph{Polymer} \textbf{2013}, \emph{54}, 320--329\relax
\mciteBstWouldAddEndPuncttrue
\mciteSetBstMidEndSepPunct{\mcitedefaultmidpunct}
{\mcitedefaultendpunct}{\mcitedefaultseppunct}\relax
\EndOfBibitem
\bibitem[Arruda and Boyce(1993)Arruda, and Boyce]{Arruda1993}
Arruda,~E.~M.; Boyce,~M.~C. A Three-Dimensional Constitutive Model for the Large Stretch Behavior of Rubber Elastic Materials. \emph{J. Mech. Phys. Solids} \textbf{1993}, \emph{41}, 389--412\relax
\mciteBstWouldAddEndPuncttrue
\mciteSetBstMidEndSepPunct{\mcitedefaultmidpunct}
{\mcitedefaultendpunct}{\mcitedefaultseppunct}\relax
\EndOfBibitem
\bibitem[Qi \latin{et~al.}(2009)Qi, \latin{et~al.} others]{Qi2009}
Qi,~Y.; others Nanostructure Effects on the Mechanical Behavior of Polymer Composites. \emph{Nat. Mater.} \textbf{2009}, \emph{8}, 882--887\relax
\mciteBstWouldAddEndPuncttrue
\mciteSetBstMidEndSepPunct{\mcitedefaultmidpunct}
{\mcitedefaultendpunct}{\mcitedefaultseppunct}\relax
\EndOfBibitem
\end{mcitethebibliography}

\end{document}